\documentclass[twocolumn, amssymb, nobibnotes, aps, prb, superscriptaddress]{revtex4-2}

\newcommand{\kk}{_{\vec k}}
\newcommand{\be}{\begin{equation}}
\newcommand{\ee}{\end{equation}}

\usepackage{amsmath, amsfonts, amstext, amssymb}
\usepackage{siunitx}
\usepackage{exscale}
\usepackage{ae}
\usepackage{graphicx, wrapfig}
\usepackage{enumerate}
\usepackage{wrapfig}
\usepackage{placeins}
\usepackage[dvipsnames]{xcolor}

\usepackage[normalem]{ulem}

\begin{document}

\title{Lattice-driven femtosecond magnon dynamics in $\alpha$-MnTe}%

\author{Kira Deltenre}
\affiliation{Condensed Matter Theory, Department of Physics, TU Dortmund University,
44227 Dortmund, Germany}
\author{Davide Bossini}
\affiliation{Department of Physics and Center for Applied Photonics, University of
Konstanz, D-78457 Konstanz, Germany}

\author{Frithjof B. Anders}
\affiliation{Condensed Matter Theory, Department of Physics, TU Dortmund University,
44227 Dortmund, Germany}
\author{G\"otz S. Uhrig}
\affiliation{Condensed Matter Theory, Department of Physics, TU Dortmund University,
44227 Dortmund, Germany}
\date{\today}%

 \begin{abstract}
The light-induced femtosecond dynamics of the sublattice magnetizations
in the  antiferromagnetically ordered phase of the semiconductor $\alpha$-MnTe is
investigated theoretically as function of an
external driving field. The electromagnetic field is coupled to optical
modes and the concomitant atomic displacements modulate the
Heisenberg exchange couplings.
We derive the equations of motion for the time-dependent sublattice magnetization in
spin wave theory and analyze the contributions from the driven magnon modes.
The antiferromagnetic order parameter exhibits coherent longitudinal
oscillations determined by the external
driving frequency which  decay due to dephasing. Including a phenomenological dissipative term to mimic spin-lattice relaxation processes leads to relaxation back to thermal equilibrium.
We provide approximate analytic solutions of the resulting
differential equations which allow us to understand the effect
of the driving light pulse on the amplitude, frequency, and lifetime of the
coherent spin dynamics.
%
 \end{abstract}

\maketitle

\section{Introduction}

The interaction between light and matter is one of the central current
issues of condensed matter physics due to the various types of magnetic
and electronic order and different essential interactions in materials.
Besides the general understanding of
light-matter interactions, specifically, controlling the magnetic order of solids
with light \cite{RevModPhys.82.2731, kimel2005ultrafast, PhysRevLett.76.4250,
stupakiewicz2017ultrafast, koopmans2010explaining, Hofherreaay8717} could be the
required milestone to develop spintronic devices
working on unprecedented timescales. In experiments,
femtosecond laser pulses have already realised the ultrafast all-optical switching of the magnetisation in a wide variety of materials
\cite{RevModPhys.82.2731, PhysRevLett.76.4250}, femtosecond magnetic phase transitions
\cite{PhysRevLett.116.097401,PhysRevLett.108.037203,bossi18},
and the coherent generation of magnons
\cite{kimel2005ultrafast,bossini2016macrospin,PhysRevB.89.060405,SatoEtAl2015},
even with the mediation of the lattice \cite{NovaEtAl2017}.
The latter effect is particularly relevant for more reasons:
(i) coherent mechanisms provide the possibility to manipulate spins without energy dissipation; (ii) the magnetoelastic coupling is almost ubiquitous
and significantly strong in antiferromagnets \cite{NemecEtAtl2018,gomon21}
which are highly promising and heavily investigated compounds.

Here, we  focus on a specific strongly correlated compound, namely hexagonal manganese telluride ($\alpha$-MnTe) \cite{Mobasser}.
It is a magnetic semiconductor with indirect bandgap in the near infrared range ($E_g = (\num{1.27}-\num{1.46})\,\si{\electronvolt}$ \cite{Kriegner_2016}).
While the semiconducting properties arise from
the Te $5p$ orbitals  and the Mn $4s$ orbitals, the material
exhibits
an antiferromagnetic (AF) order of the Mn $3d$ spins consisting of ferromagnetically ordered layers of spins which are antiparallel between adjacent layers  below the N\'{e}el temperature
$T_\text{N} \approx \SI{310}{\kelvin}$ \cite{Kriegner_2016}.
It is found that the electronic band gap is influenced by the
degree of magnetic order. In fact, an additional contribution to the band gap
proportional to the square of the sublattice magnetization occurs
\cite{PhysRevB.61.13679,Bossini_2020}.

This material has been experimentally proven to possess a substantial
spin-lattice coupling, since both the frequency and the lifetime of
two degenerate Raman-active phonon modes with $5.3$ THz frequency are
significantly affected by the establishment of the long-range magnetic
order \cite{RamanMnTe2020}.
Hexagonal MnTe is thus a representative choice for the material class of dielectric correlated antiferromagnets. The correlated nature is in fact common to a massive variety of other compounds (e.g. oxides) and the strong magneto-acoustic coupling is almost ubiquitous in antiferromagnets \cite{RamanMnTe2020,gomon21}.
So far two phonon-magnon coupling mechanisms have been considered to
interpret observations on the femtosecond time-scale:
the nonlinear phononics \cite{FoerstEtAl2011,NovaEtAl2017}
and the Kittel mode \cite{PhysRevB.97.140404}.

A very recent time-resolved experiment \cite{bossiniMnTeExp2021}
displays photo-induced coherent oscillations of the rotation of the polarisation of light well pronounced
in the antiferromagnetic phase of $\alpha$-MnTe. In this experiment,
femtosecond laser pulses trigger the $5.3$ THz modes by means of
a Raman-scattering mechanism changing the interatomic potentials
for fractions of a picosecond. This displacive effect
constitutes the pumping mechanism. Moreover, the dispersion of magnons of the material
\cite{PhysRevMaterials.3.025403}
 excludes the physical framework of the Kittel mode because magnons and phonons do not display an avoided crossing in their dispersions.
  A novel physical mechanism has thus to be explored to interpret the observations, which are expected
not to be limited to $\alpha$-MnTe, given the generality of the formulation.

The question arises how light couples to the magnetic subsystem and
coherently drives a magnetization modulation in the class of
antiferromagnetically ordered insulators (or semiconductors)
\cite{bossiniMnTeExp2021}.
Based on the seminal work by Fleury and Loudon \cite{FleuryLoudon1968}
and Shastry and Shraiman \cite{ShastryShraiman1990} it is
possible that Raman scattering directly excites spin degrees of freedom
via the Peierls coupling of the vector potential in the hopping
elements of the underlying fermionic Hubbard model.
This mechanism has been recently invoked to
compute the magnon circular photogalvanic effect \cite{Bostroem2021}.
The mapping of the fermionic model on the spin model works
best if the Raman pumping is performed
off resonance. But
in this regime the excitation of magnetic dynamics tends
to be rather inefficient.

In contrast to the direct coupling of light
to the spins via a Peierls substitution in the underlying
Hubbard model
\cite{ShastryShraiman1990}, the aforementioned dominant coupling of optical phonons to the spin degrees of freedom,
is revealed by Raman
spectroscopy \cite{RamanMnTe2020}. This coupling can be
qualitatively understood  since the superexchange paths pertinent to the magnetic couplings depend on the overlaps of the
atomic orbitals, which are determined by the ionic positions and possible motions thereof.
This is a key ingredient for creating magnetic excitations.

Such atomic motion can be triggered by so called displacive
stimulated Raman scattering. The excitation
of an electron from an occupied to an unoccupied state alters
the interionic potentials so that the previous atomic positions
are no longer equilibrium positions. Hence the atoms start moving
towards the minima of the modified potential.
This induces a coherent oscillatory motion of the
atomic positions in each unit cell that is governed  by
the optical phonon  frequencies.
This oscillatory motion persists once the electronic
excitations has decayed on the short time scale of
hundreds of femtoseconds.
This periodic modulation of the Mn-Te distances and angles in the unit cell
translates into a periodic modulation of the tight-binding hopping
parameters of an effective Hubbard model for the manganese $3d$ subsystem.
According to the relation $J\propto t^2/U$ between the magnetic
exchange $J$, the hopping $t$ and the on-site repulsion $U$
the periodic modulation of $t$ entails a periodic modulation of the
exchange coupling \cite{PhysRevMaterials.2.064401}.

%

This leads us to a coupling mechanism
by means of displacement
based on the modulation of the
Heisenberg exchange couplings in an effective spin model
for the magnetic degrees of freedom. This optomagnetic mechanism
is similar to the one recently studied for a
disordered spin system \cite{PhysRevB.103.045132}.
The key difference is the way how
the phonons or atomic displacements
are pumped by light. Infrared active phonons
can  be set into motion directly by THz radiation
while we invoke displacive stimulated Raman scattering
in the present work.

A indirect magneto-phononic coupling was
considered in the context of Cr$_2$O$_3$.
The assumption was made in Ref.\ \cite{PhysRevMaterials.2.064401}
that an infrared  vibrational model  is driven
by sinusoidal laser field in the THZ range,
and is coupled via non-linear terms to an effective Heisenberg model by a Raman mode. The magnetization dynamics was calculated
by a phenomenolgical  Landau-Lifshitz-Gilbert equation.

In the present paper, we are guided by the experimental setup \cite{bossiniMnTeExp2021} where a very short fs laser pulse with spectral range centred at 1.71 eV  slightly above the band gap
induces atomic displacements on an ultrashort time scale
leading to a coherent optical lattice mode
which drives the magnetic subsystem on a ps time scale.
The amplitude of the lattice mode decays exponentially on the ps time scale due to coupling to the bath of acoustic phonons.
In particular, we are interested in the quantum mechanical description of the
change in the sublattice magnetization of the antiferromagnet as function of time.

Starting from the modulated Heisenberg exchange couplings \cite{PhysRevMaterials.2.064401},
we employ Holstein-Primakoff linear spin-wave theory
out of equilibrium to calculate the generation
of magnon pairs and the induced dynamics of the
sublattice magnetization for various system parameters
such as the driving frequency, the driving intensity, and
the pulse duration. We also include spin-lattice relaxation
on the level of a Markov approximation
to describe the realistic long-time relaxation of the
optically activated system.

Our approach is generic for lattice-driven
magnon dynamics in ordered quantum magnets
leading to a universal analytic structure
of the equations of motion. The
material specific details only enter in the modifications
of momentum dependent couplings in the differential equations,
but they do not change their analytic structure.
Using realistic
exchange couplings extracted from measured magnon dispersions  \cite{PhysRevB.73.104403}
allows us to make contact to the lattice-driven femtosecond magnon dynamics in $\alpha$-MnTe  \cite{bossiniMnTeExp2021}.

This article is structured as follows.
We present the underlying Heisenberg spin model in
Sec.\ \ref{sec:model}
and its treatment by Holstein spin-wave theory in
Sec.\ \ref{sec:spin-wave-theory}
where we establish the equilibrium properties in the
antiferromagnetic symmetry broken phase.
Section \ref{sec:NEQ} is devoted to the non-equilibrium dynamics.
We derive the generic equation of motion for the driven magnon dynamics in Sec.\ \ref{sec:driven-magnons}
and discuss the properties in Sec.~\ref{sec:specific_modes} where we present
approximate analytic solutions of the equations. The accuracy of the
approximations is demonstrated by a comparison with the numerical solution of the full dynamics.
Section \ref{sec:pulse-magnetization} is devoted to the various aspects of the dynamics
of the sublattice magnetization as induced by the pulses.
We use the developed theoretical framework to make
contact to the magnon dynamics in $\alpha$-MnTe in
Sec.~\ref{sec:contact-to-experiments}.
The paper is finished with conclusion and outlook in
Sec.\ \ref{sec:conclusion}.

\section{Theory}

The focus of our work are the collective magnetic excitations of an
antiferromagnet and, in particular, their coupling to the lattice. Therefore, we
focus on the purely magnetic description based on an effective Heisenberg model.

\subsection{Model}
\label{sec:model}

\begin{figure}[t]
  \centering
  \includegraphics[width=0.98\columnwidth,clip]{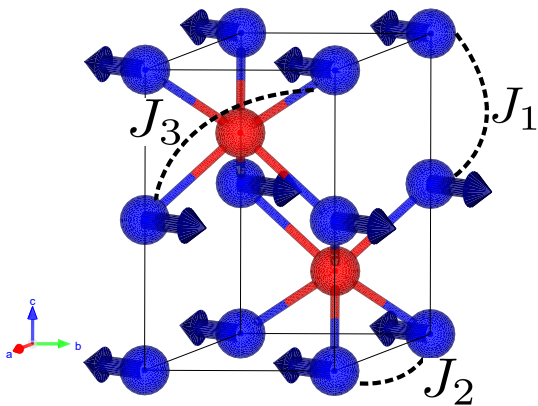}
  \caption{The unit cell of $\alpha$-MnTe consists of two Mn-ions (blue) and two Te-ions (red).
        On the Mn-sites, magnetic moments are localized. Image made with VESTA \cite{VESTA}.
          }
  \label{fig:unitcell}
\end{figure}

The unit cell of hexagonal MnTe with NiAs structure consists of two Mn-ions and
two Te-ions as depicted in Fig.~\ref{fig:unitcell}. At each Mn-site,
a magnetic moment of $S=5/2$ is localized. The Mn-sites are arranged in layers of
stacked triangular lattices. The spins within each triangular layer are aligned
in a parallel way.
The spins of the adjacent layers are antiparallel to each other as indicated by the arrows
in Fig.~\ref{fig:unitcell}. The two Mn-ions in the unit cell belong to different
sublattices
which are defined by the orientation of the spins.
The model captures the magnetic moments localized at the Mn-sites.
The orbitals of the Te-ions allow for virtual hopping processes generating
exchange couplings. They do not need to be considered explicitly here.

The hybridization between the Mn-Mn and Mn-Te orbitals mediate effective Heisenberg
couplings between the localized $S=5/2$ Mn $3d$ moments. The resulting low-temperature
spin Hamiltonian reads
\begin{eqnarray}
    H &= & J_1 \sum_{\langle i, j \rangle_c} \vec{S}_i \vec{S}_j + J_2
                \sum_{\langle i, j \rangle_{ab}} \vec{S}_i \vec{S}_j
+ J_3 \sum_{\langle \langle i, j \rangle \rangle} \vec{S}_i \vec{S}_j
  \label{eq:hamiltonian}
\end{eqnarray}
as derived in  Refs.~\cite{PhysRevB.73.104403,PhysRevMaterials.3.025403}.
The model includes three terms: an antiferromagnetic interlayer coupling $J_1>0$
between nearest neighbors (NN) along the $c$-axis, a ferromagnetic in-plane coupling
$J_2<0$, and an antiferromagnetic third-nearest-neighboring coupling $J_3>0$.
In Ref.~\cite{PhysRevB.73.104403} the values for the parameters were obtained by a fit
to inelastic neutron scattering data based on spin-wave theory leading to the values
$J_1= \SI{21.5}{\kelvin}, J_2=\SI{-0.67}{\kelvin}$, and $J_3= \SI{2.87}{\kelvin}$.
Due to the different notation of the Hamiltonian, the signs are different
in this work, and a factor 2 is included.

The interaction between femtosecond laser pulses and MnTe is expected to induce coherently
Raman active phonons, either via the impulsive stimulated Raman scattering
\cite{MERLIN1997207}  or displacive excitation of coherent phonons mechanism \cite{Zeiger1992}.
In particular, the 5.3 THz modes aforementioned are optical phonons and they modify the exchange coupling, as they correspond to Te-atoms oscillations \cite{bossiniMnTeExp2021}.
 We consider the resonant excitation of a particular phonon
at frequency $\omega_0$. Hence, this is also the frequency by which the
the exchange couplings are modulated.
We assume that the parameter $J_3$, which couples third-nearest-neighboring spins
via a Te-ion, changes its value due to the relative oscillation of the Mn- and the
Te-ions.
We include this effect in the Hamiltonian by a time-dependent coupling
\begin{equation}
J_3\to J_3(t)= J_3^{(0)} +\delta J_3(t)
\end{equation}
 where $J_3^0$ is the equilibrium exchange coupling
and $\delta J_3(t)$ parametrizes the effect of the laser field via the phonon
on the magnetic subsystem.
In general, the other exchange parameters can also change in time.
But for clarity, we stick to the modulation of $J_3$ for the majority of calculations.
The modulation of the other couplings is considered in Sec.~\ref{sec:with_other_deltaJ}
and it turns out that the effect of modulated $J_1$ or $J_2$ is qualitatively very
similar
and does not lead to qualitatively different phenomena.

\subsection{Method}
\label{sec:spin-wave-theory}

In order to analyze the non-equilibrium dynamics of the order parameter,
we first determine the equilibrium model by resorting to a
linear spin waves ansatz \cite{PhysRevB.40.2494,PhysRevB.46.6276}
for the AF phase.
As a first step, we invert all spins of one sublattice such that the mapped
problem has ferromagnetic spin order. In a second step,
we employ the Holstein-Primakoff representation \cite{PhysRev.58.1098} which
represents the spin operators by bosonic operators
\begin{subequations}
\begin{align}
  S_i^z &= -S + \tilde{b}^\dagger_i \tilde{b}_i
        \\
  S_i^+ &= \tilde{b}_i^\dagger \sqrt{2S - \tilde{b}^\dagger_i \tilde{b}_i}
        \approx \sqrt{2S} \tilde{b}_i^\dagger
        \\
  S_i^- &= \sqrt{2S - \tilde{b}_i^\dagger \tilde{b}_i} \tilde{b}_i
        \approx \sqrt{2S}
        \tilde{b}_i
\end{align}
\end{subequations}
keeping only the leading order in a $1/S$ expansion. This an excellent approximation
in the AFM phase because of the large spin $S=5/2$ and the large number of coupled
neighboring spins.
The $z$-direction in spin space is defined by the orientation of the spins of
sublattice A in the ordered phase. Note that $z$-axis of the spin lies in fact
in the $ab$-plane in real space and that we choose $\tilde{b}$ for the bosonic
operators at the lattice sites and will use $b$ later in the diagonal
Bogoliubov representation.

Note that we only have to deal with bilinear terms in leading order.
Thereby, interactions between different $\vec{k}$-modes are neglected;
in this approximation they do not scatter from each other.
In addition, relaxation mechanisms are not included in our model so far.
The Fourier transform
\begin{equation}
  \tilde{b}_i = \frac{1}{\sqrt{N}} \sum_{\vec{k}}
        \exp\left(i \vec{k}\cdot \vec{l}\right) \tilde{b}_{\vec{k}},
\end{equation}
into momentum space yields the dimensionless Hamiltonian
\begin{equation}
  \frac{H_0}{J_1 S} =
        E_d + \sum_{\vec{k}} \Big[ A_{\vec{k}} \tilde{b}_{\vec{k}}^\dagger
\tilde{b}_{\vec{k}}
        + \frac{1}{2} B_{\vec{k}} \left(\tilde{b}_{\vec{k}}^\dagger
\tilde{b}_{-\vec{k}}^\dagger
        + \text{h.c.}\right) \Big]
        \label{eq:hamiltonian_h0}
\end{equation}
where we use the coefficients
\begin{subequations}
\begin{align}
        A_{\vec{k}} &:= \frac{1}{J_1} \Big(2 J_1 - 6J_2 + 12J_3
    + J_2\gamma_\Delta (\vec{k})\Big)
                \\
  B_{\vec{k}} &:= 2 \cos(k_c) \Big( 1+ 2\frac{J_3}{J_1} \gamma_\Delta \left(\vec{k}\right)\Big)
        \label{eq:B_def}
        \\
        \nonumber
 \gamma_\Delta\left(\vec{k}\right)
                &:= \cos(k_a) + \cos \left(\sqrt{3}k_b/2 +  k_a/2 \right)
                \\
     &\qquad + \cos\left(-\sqrt{3}k_b/2 + k_a/2 \right).
\end{align}
\end{subequations}
The Hamiltonian is diagonal in $k$-space except for the coupling of
$\vec k$ and $-\vec k$ in the creation
and annihilation terms of pairs of magnons.
his also determines how non-equilibrium perturbations in the
couplings
enter the real-time dynamics investigated below.

The Holstein-Primakoff transformation generates the offset $E_d$
independent of momentum
\begin{equation}
  E_d = -NS \left(1 -3 \frac{J_2}{J_1} + 6 \frac{J_3}{J_1}\right).
\end{equation}
The Bogoliubov transform
\begin{equation}
  \tilde{b}_{\vec{k}} = b_{\vec{k}} \cosh{\theta_{\vec{k}}} + b_{-\vec{k}}^\dag
\sinh{\theta_{\vec{k}}}
\end{equation}
fully diagonalizes the Hamiltonian \eqref{eq:hamiltonian_h0}
\begin{equation}
  \frac{H_0}{J_1 S} =
  \sum_{\vec{k}} \omega_{\vec{k}} b_{\vec{k}}^\dagger b_{\vec{k}} + E_d + \Delta E
  \label{eq:H0-diagonal}
\end{equation}
if the condition for the Bogoliubov angle $\theta_{\vec{k}}$
\begin{equation}
  \frac{B_{\vec{k}}}{A_{\vec{k}}} = -\tanh(\theta_{\vec{k}})
\end{equation}
is met. The additional  contribution $\Delta E$ to  the ground energy is given by
\begin{equation}
  \Delta E = \frac{1}{2N} \sum_{\vec{k}} \left(\omega_{\vec{k}} - A_{\vec{k}} \right).
\end{equation}
It does not influence the magnetic excitations.
Finally, we obtain the  magnon dispersion in units of $J_1$
\begin{equation}
   \omega_{\vec{k}} = \sqrt{A_{\vec{k}}^2 - B_{\vec{k}}^2}\,,
  \label{eq:dispersion}
\end{equation}
which is easily evaluated for any given set of couplings $J_i$.

\subsubsection{Equilibrium properties}

\begin{figure}[t]
  \includegraphics[width=\columnwidth]{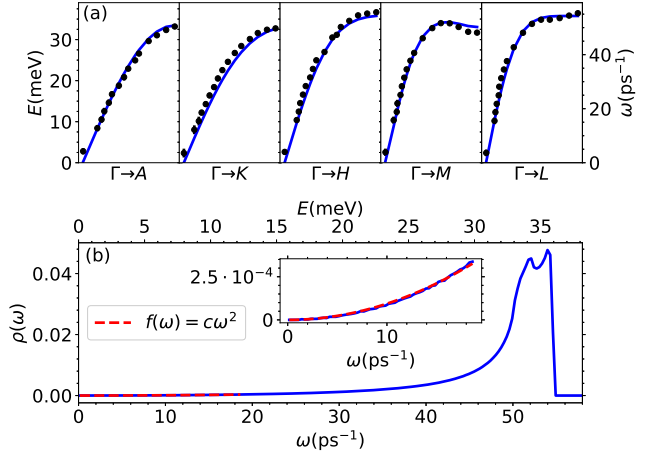}
  \caption{The magnon dispersion (a) and the magnon density-of-states (DOS) (b).
        (a) The magnon dispersion measured by inelastic neutron scattering (black dots
        from Ref.~\cite{PhysRevB.73.104403}) can be described well by the parameter set
        proposed in that reference. (b) The corresponding magnon DOS with the  parabolic onset
        (detailed illustration in inset) established for a linear dispersion at low
energies in
        three dimensions. The van Hove singularities appear close to the
        the maximum energy $\hbar\omega_\text{max}$, i.e., most energies lie close to
        $\hbar\omega_\text{max}$.
        }
  \label{fig:dispersion_DOS}
\end{figure}

The parameter set stated below Eq.\ \eqref{eq:hamiltonian}
yields the magnon dispersion shown as blue continuous curve
in Fig.~\ref{fig:dispersion_DOS}(a) in various directions in the Brillouin zone.
We use it  for the following calculations.
For comparison, the experimental data from Ref.~\cite{PhysRevB.73.104403} are
included as well as black filled circles.

In the calculation of the magnon density-of-states (DOS)
\begin{equation}
  \rho (\omega) = \frac{1}{N} \sum_{\vec{k}} \delta (\omega - \omega_{\vec{k}} ).
\end{equation}
for the stacked triangular layers, we profit from the analytical results for
triangular lattices in Ref.~\cite{doi:10.1002/andp.19955070405}. The additional
vertical component contributes only the term $\cos(k_c)$ in Eq.~\eqref{eq:B_def}.
The resulting magnon DOS shown in Fig.~\ref{fig:dispersion_DOS}(b) sets in
quadratically
\begin{equation}
  f(\omega) = c \omega^2
\end{equation}
with $c=\SI{ 9.40(4)e-7}{\pico\second}$ (red dashed line)
as implied by the linear dispersion and the three dimensions.
The maximum energy $\hbar\omega_\text{max}\approx \SI{35.9}{\milli\electronvolt}$
is given by the maximum of the magnon dispersion.
In the density of states, we see two peaks that are related to van Hove singularities.
The second van Hove singularity is almost at the maximum frequency. Essentially,
most frequencies lie close to the maximum frequency.

The order parameter of the antiferromagnetic phase is given by the sublattice
magnetization per spin defined as
\begin{align}
  L &= \frac{1}{N} \sum_i (-1)^{\delta_i} \langle \hat{S}_i^z \rangle
\\
  \delta_i &=
  \begin{cases}
    0 & \text{ for sublattice A}\\ \nonumber
    1 & \text{ for sublattice B.} \nonumber
  \end{cases}
\end{align}
We express the order parameter in the framework of
 spin-wave theory by applying the same
transformations as outlined above leading to
$L=L_0 -\delta L$. The first term $L_0$ is temperature independent and
determined by the spin size $S=5/2$ and the quantum fluctuations
\begin{subequations}
\begin{align}
L_0 &=  S - \Delta S \\
  \Delta S &= \frac{1}{2N}
        \sum_{\vec{k}} \left (\frac{\omega_{\vec{k}}}{A_{\vec{k}}^2} -1 \right)\,.
\end{align}
\end{subequations}
The temperature dependence enters in the second term
\begin{eqnarray}
  \delta L&=&
  \frac{1}{N}
  \sum_{\vec{k}} \left[\frac{A_{\vec{k}}}{\omega_{\vec{k}}}
        \langle n_{\vec{k}} \rangle - \frac{B_{\vec{k}}}{\omega_{\vec{k}}}
        \Re \langle b_{\vec{k}}^\dagger b_{-\vec{k}}^\dagger \rangle \right]
        \label{eq:delta-L-def}
\end{eqnarray}
The second term in the square bracket
of Eq.~\eqref{eq:delta-L-def} allows us to treat also non-equilibrium situations.

In the case of a driving term in the Hamiltonian, $\delta L$
becomes time dependent.
Due to the upper bound of $L=5/2$, the model only yields reliable results
if $|\delta L|\ll L$. Thus, the change of the sublattice magnetization
$\delta L$ has to stay significantly
 smaller than the constant part $L_0$ in order for the spin wave theory
to be applicable. In this paper, we focus on the effect of driving on
the sublattice magnetization deep in the ordered phase. We
do not consider the effects of finite temperature so that we
assume $T=0$.

\FloatBarrier

\section{Dynamics out of equilibrium}
\label{sec:NEQ}

\subsection{Magnon dynamics driven by pulses}
\label{sec:driven-magnons}

To describe the dynamics out of equilibrium, we consider
the following Hamiltonian \cite{PhysRevMaterials.2.064401}
including driving
\begin{equation}
  H(t) = H_0 + X(t)\,
  \label{eq:energy_timedependent}
\end{equation}
which consists of the Hamiltonian $H_0$
of the system in equilibrium  and of the  time dependent driving
$X(t)$.
The Hamiltonian $H_0$ and $H(t)$ are both given by
the original form of Hamiltonian,  Eq.\ \eqref{eq:hamiltonian}, and only differ
in the parameters $J_i$. While $H_0$ only contains the equilibrium
values $J_i^{(0)}$, the Heisenberg couplings
in $H(t)$ are supplemented by a time-dependent part, $J_i(t) =  J_i^{(0)} +\delta J_i(t)$.

We include the effect of the laser pulse on
the magnetic subsystem of MnTe by time dependent oscillations
$\delta J_{3}(t)$.
As already mentioned, the laser pulses excite an optical
phonon which in turn modifies the exchange path and thereby
the exchange coupling \cite{PhysRevB.103.045132}.
Therefore, we replace $J_3\to J_3^{(0)} + \delta J_3(t)$ in the
last term in Eq.\ \eqref{eq:hamiltonian} and use the resulting
term proportional to  $\delta J_3(t)$ as driving operator $X(t)$.

After the Bogoliubov transformation it takes the form
\begin{align}
\label{eqn:J3-driving}
 X(t) &= S \delta J_3(t) \sum_{\vec{k}}
\Big[ \alpha_{\vec{k}} b_{\vec{k}}^\dagger b_{\vec{k}} +
\frac{1}{2} \beta_{\vec{k}}
\left(b_{\vec{k}}^\dagger b_{\vec{k}}^\dagger
+ \text{h.c.} \right) + C_{\vec{k}} \Big]
\end{align}
with the $\vec{k}$-dependent coefficients
\begin{subequations}
\label{eq:alpha_beta}
\begin{align}
\alpha_{\vec{k}} &= \frac{A_{\vec{k}}}{\omega_{\vec{k}}}
\left( 12 - 4\frac{B_{\vec{k}}}{A_{\vec{k}}} \cos(k_c) \gamma_\triangle
(\vec{k}) \right)
\\
  \beta_{\vec{k}} &= \frac{A_{\vec{k}}}{\omega_{\vec{k}}}
        \left( -12 \frac{B_{\vec{k}}}{A_{\vec{k}}} + 4 \cos(k_c)
        \gamma_\triangle (\vec{k}) \right).
\end{align}
\end{subequations}
Note that only  equilibrium parameters $J_i^{(0)}$ enter
the Bogoliubov coefficients.

The driving in Eq.\ \eqref{eqn:J3-driving} includes the crucial
term proportional to $ \beta_{\vec{k}}$ creating two magnons.
This term is not only the source for incrementing
the magnon occupation and other amplitudes,
but its also the reason for a frequency doubling
in the resonance condition which will be derived below.

Perturbing only the coupling $J_3$ as a consequence of the excitation of the
phonon is one choice among others. We motivate it by the fact
that its exchange path runs through the Te-ions so that it
is mostly affected by the relative motion of Mn- and Te-ions
generated by the lattice dynamics.
Nevertheless, we cannot exclude a priori lattice-driven modulations of the other exchange couplings
$J_1$ or $J_2$. Their influence is subject of
Sec.\ \ref{sec:with_other_deltaJ} below.
It turns out that only the parameters $\alpha_{\vec{k}}$ and
$\beta_{\vec{k}}$ need to be changed to account for the
modulation of either $J_1$ or $J_2$.
But the analytic structure of the differential equations remains
unaltered since all interactions are described by a Heisenberg Hamiltonian.
The key differences are the coupling strengths $J_i$ entering in the
relative modulation amplitude
$a_0$ introduced below in
Eq.~\eqref{eq:DJ3-vs-t}) and the number and positions
of nearest neighbors spins defining
$\alpha_{\vec{k}}$ and $ \beta_{\vec{k}}$.

The  coefficient $C_{\vec{k}}$ in Eq.\ \eqref{eqn:J3-driving}
only changes the ground state energy and does not
create or annihilate any magnons. For this reason, it is omitted
in the following.
The remaining part of the operator $X(t)$ is non-diagonal in the
Bogoliubov bosons and drives magnon excitations, i.e.,
creates or generates them or changes their energy.

The premiss of our work is that an ultrashort laser fs
pulse with spectral range centered at \SI{1.71}{\electronvolt}
provides a displacement of the charged Mn and Te ions
that happens on a fs time scale. The induced
damped coherent lattice vibrations characterized by
the optical phonon frequencies
causes a modulation $\delta J_i(t)$ of Heisenberg
coupling constants due to the periodic change of the hopping matrix elements.
From this mechanism we can conclude that the pulse shape of the optical pulse
does not enter the driving term, and the optical fs pulse can be treated as instantaneous
compared to the phonon and magnon frequencies which are in the small THZ range.

We parameterize the effect of this phonon on the magnetic
system by damped oscillations of $\delta J_i(t)$. They read in
their dimensionless form
\begin{align}
\label{eq:DJ3-vs-t}
   \frac{\delta J_i(t)}{S J_1^{(0)}} &= a_i(t) = a_{0,i}
        \exp(-\gamma t) \cos(\omega_0 t),
\end{align}
where $a_{0,i}= \delta J_i(0)/J_1^{(0)} = (J_i/J_1^{(0)}) \kappa_i$.
In the last step, we introduce the relative change
\begin{equation}
  \kappa_i =\delta J_i(0)/J_i^{(0)}
  \label{eq:kappa_i}
\end{equation}
while the ratio $(J_i/J_1^{(0)})$ links $a_{0,i}$ to the parameters
of the equilibrium Hamiltonian.
Note that the driving field for the magnon dynamics is governed by the
coherent lattice oscillations  \cite{PhysRevMaterials.2.064401}
and not by the time scale of the optical laser pulse.

Once the equilibrium parameters are fixed,
the perturbation amplitude is uniquely parameterized by $\kappa_i$.
The assumption that expression \eqref{eq:DJ3-vs-t} describes the
modulation of the exchange coupling is based on the results of
previous calculations for a related model in which the phonon dynamics
has been computed explicitly \cite{PhysRevB.103.045132}.  Of course, Eq.~
\eqref{eq:DJ3-vs-t} represents a simplification, but it is a
reasonable one for small spin-phonon coupling. Moreover, it
contains two main ingredients: the amplitude $a_{0,2}$
 and the relaxation rate $\gamma$ of the modulation.
In the following, we focus on the effect of $\delta J_3(t)$ and omit the
index $i$ therefore. In Sec.~\ref{sec:with_other_deltaJ}, we come back
to modulations of $\delta J_1(t)$ and $\delta J_2(t)$.

Describing the oscillations of the coupling by $\cos(\omega_0 t)$ is
consistent with the displacive excitation mechanism of the lattice modes \cite{MERLIN1997207}.
We access the time dependent sublattice magnetization $L(t)$ in
Eq.~\eqref{eq:delta-L-def} by the expectation values of the magnon occupation
\begin{subequations}
  \label{eq:uvw_k}
  \begin{align}
   u_{\vec{k}} &:= \langle n_{\vec{k}} \rangle = \langle b_{\vec{k}}^\dagger
b_{\vec{k}} \rangle
        \label{eq:u_k}
        \\
        \intertext{and of the real and imaginary part of
        the off-diagonal term $\langle b_{\vec{k}}^\dagger b_{-\vec{k}}^\dagger \rangle $ }
   v_{\vec{k}} &:= \text{Re}\, \langle b_{\vec{k}}^\dagger b_{-\vec{k}}^\dagger
\rangle\\
   w_{\vec{k}} &:= \text{Im}\, \langle b_{\vec{k}}^\dagger b_{-\vec{k}}^\dagger \rangle
        \label{eq:w_k} .
  \end{align}
\end{subequations}
whose dynamics are calculated by means of
Heisenberg's equation $\frac{dA}{dt} = i \langle [H(t),A] \rangle$.
Since the Hamiltonian does not break translational invariance even in
presence of the driving the resulting differential equations
\begin{subequations}
\label{eq:eom_uvw}
\begin{eqnarray}
 \frac{d u_{\vec{k}}}{dt} &= &2 a(t) \beta_{\vec{k}} w_{\vec{k}} -\gamma_r u_{\vec{k}}
        \label{eq:eom_u}
        \\
  \frac{d v_{\vec{k}}}{dt} &=& -2(\omega_{\vec{k}} + a(t) \alpha_{\vec{k}}) w_{\vec{k}}
        \label{eq:eom_v} -\gamma_r v_{\vec{k}} \\
  \frac{d w_{\vec{k}}}{dt} &= &2(\omega_{\vec{k}} + a(t) \alpha_{\vec{k}}) v_{\vec{k}}
        +2 a(t)  \beta_{\vec{k}} (u_{\vec{k}} +1/2)\nonumber \\
        &&  -\gamma_r w_{\vec{k}}
  \label{eq:eom_w} ,
\end{eqnarray}
\end{subequations}
do not couple different $\vec k$-values. But all of them
contribute to the collective sublattice magnetization
in \eqref{eq:delta-L-def}.

In addition to the unitary dynamics induced by the Hamiltonian
in the Heisenberg equations of motion we introduce a
phenomenological relaxation rate $\gamma_r$ describing
the decay of magnons. By this parameter we quantify the
coupling to thermal bath,
 i.e., we treat the
driven spin system as open quantum system and describe it
by a Lindblad equation \cite{breue06}
where the creation and annihilation operators of the magnons
serve as Lindblad operators \cite{PhysRevB.103.045132}.
For simplicity, no $\vec k$-dependence
of the relaxation rate $\gamma_r$ is taken into account.

We solve the set of differential equations for a mesh
of $M$ discrete values $k_c$ and a mesh of $M$ discrete frequencies
for the density-of-states of the triangular lattice
as computed analytically \cite{doi:10.1002/andp.19955070405}.
Note that we absorb the factor $\hbar$ into the definition of
 time $t$ as measured in units of the inverse energy $1/J_1$.
The numerical results presented in the next sections are obtained by
solving the equations~\eqref{eq:eom_u}-\eqref{eq:eom_w}
with a Bulirsch-Stoer algorithm \cite{10.5555/1403886}
with \num{1000} time steps per picosecond. We discretize the
Brillouin zone in $c$-direction. For the Mn-planes parallel
to the $ab$-plane, we use the density of states for a
triangular lattice \cite{doi:10.1002/andp.19955070405}.
Since we treat the system prior to any pumping
to be at zero temperature at equilibrium, we use the
initial conditions $u_{\vec{k}} (t=0) = v_{\vec{k}} (t=0) =
w_{\vec{k}} (t=0) = 0$.

\subsection{Dynamics of specific magnon modes}
\label{sec:specific_modes}

Before addressing the total sublattice magnetization
we analyze the  differential equations of Eq.~\eqref{eq:eom_uvw}
to understand the dynamics of the expectation values
for a single $\vec{k}$-mode.
If the driving term $a(t)$ and the relaxation
$\gamma_r=0$ are set to 0, the magnon occupation does not change
\begin{equation}
  \frac{du_{\vec{k}}(t)}{dt} = 0 \Rightarrow u_{\vec{k}}(t) = \text{const.}
\end{equation}
and $v_{\vec{k}}$ and $w_{\vec{k}}$ describe a coherent oscillation
between the real and imaginary part of the
expectation value $ \langle b_{\vec{k}}^\dagger b_{-\vec{k}}^\dagger \rangle$
\begin{subequations}
\begin{align}
  v_{\vec{k}}(t) &= c_0 \cos(2\omega_{\vec{k}} t+\varphi)\\
  w_{\vec{k}}(t) &= c_0 \sin(2\omega_{\vec{k}} t+\varphi)\,.
\end{align}
\end{subequations}
with frequency $2\omega_{\vec{k}}$. The
values of $c_0$ and $\delta$ depend on the initial conditions.
Clearly, in absence of a damping mechanism once excited
the amplitude of coherent oscillations, i.\ e.,
$|\langle b_{\vec{k}}^\dagger b_{-\vec{k}}^\dagger \rangle|$ does not change in time.
If we include a finite damping ($\gamma_r>0$) the coherent oscillation
is damped by the factor $\exp(-\gamma_r t)$.

The non-equilibrium dynamics of $\delta L$ given
in Eq.~\eqref{eq:delta-L-def} results from the
superposition of the contributions of all  $\vec{k}$-points leading to
a decay of the sublattice magnetization due to dephasing.
This dephasing results from the differing frequencies $2\omega_{\vec k}$.
The contributing range of these frequencies is $2\omega_\text{max}$
and thus $1/2\omega_\text{max}$ is an estimate for the time scale
of this dephasing.

In the presence of the driving term ($a_0\ne 0$)
the differential equations contain the source term
$a(t)\beta_{\vec{k}}$ originating from the two-magnon creation term
in the time-dependent part of the Hamiltonian
given in Eq.~\eqref{eqn:J3-driving}.
This source term drives the system away from the
fixed point $u_{\vec{k}}(t) = v_{\vec{k}}(t) = w_{\vec{k}}(t) = 0$.
The first term in Eq.\ \eqref{eqn:J3-driving} is responsible for a
modulation of the oscillation frequency by  $2a(t) \alpha_{\vec{k}}$,
i.e., $2\omega_{\vec k} \to 2\omega_{\vec k} + 2a(t) \alpha_{\vec{k}}$.
But since $\omega_{\vec k}$ dominates over $a(t) \alpha_{\vec{k}}$
this modulation is often not a sizable effect, see also below.

After the driving term $a(t)$ has essentially vanished due to
its exponential damping, the pair creation stops and
$u_{\vec{k}}(t)$ approaches a constant finite value if we neglect the
relaxation ($\gamma_r = 0$). Then the coherent
oscillations between $v_{\vec{k}}$ and $w_{\vec{k}}$
with a frequency $2\omega_{\vec{k}}$ occur for large times
$t\gg 1/\gamma$.

A detailed analytical, approximate evaluation can be found in Appendix \ref{app:analytic_approximation}.
After recombining, $z_{\vec{k}}(t) = \langle b_{\vec{k}}^\dagger b_{-\vec{k}}^\dagger \rangle
=v_{\vec{k}}(t) + i w_{\vec{k}}(t)$, we obtain
\begin{widetext}
\begin{equation}
    z_{\vec{k}}(t) = i a_0 \beta_{\vec{k}}
    \left[\frac{e^{-\gamma t}(\omega_0 \sin(\omega_0 t)
                -(2i\omega_{\vec{k}}+\gamma-\gamma_r)\cos(\omega_0 t))}
                {\omega_0^2-(2\omega_{\vec{k}} - i (\gamma-\gamma_r))^2}
    +\frac{(2i\omega_{\vec{k}} + \gamma- \gamma_r)
                e^{2i\omega_{\vec{k}} t} e^{-\gamma_r t}}
                {\omega_0^2-(2\omega_{\vec{k}} - i (\gamma-\gamma_r))^2}
    \right]
  \label{eq:z_final_solution_gamma_r}
\end{equation}
\end{widetext}
which enters Eq.\ \eqref{eq:eom_u}.

Resonantly driven modes, i.e.,
modes for which $\omega_0^2\approx (2\omega_{\vec{k}})^2$ holds,
acquire a large amplitude and contribute the most to $\delta L$.
The first term on the right hand side of
Eq.~\eqref{eq:z_final_solution_gamma_r} describes the coherent
oscillations with driving frequency $\omega_0$ and an
exponential envelope decreasing with the rate $\gamma$.
The second term includes the oscillations with twice the eigen frequency
$\omega_{\vec{k}}$ of each $\vec{k}$-mode. The factor of two stems
from the fact that magnons are created in pairs with total momentum zero and
 at $\vec k$ and $-\vec k$,
both of which have the same frequency
$\omega_{\vec{k}}= \omega_{-\vec{k}}$.
In addition, the second term decays exponentially with decay rate
$\gamma_r$. For zero relaxation of the magnon, $\gamma_r = 0$, the
amplitude of the second term is constant.

\begin{figure}[tb]
  \centering
  \includegraphics[width=1.08\columnwidth]{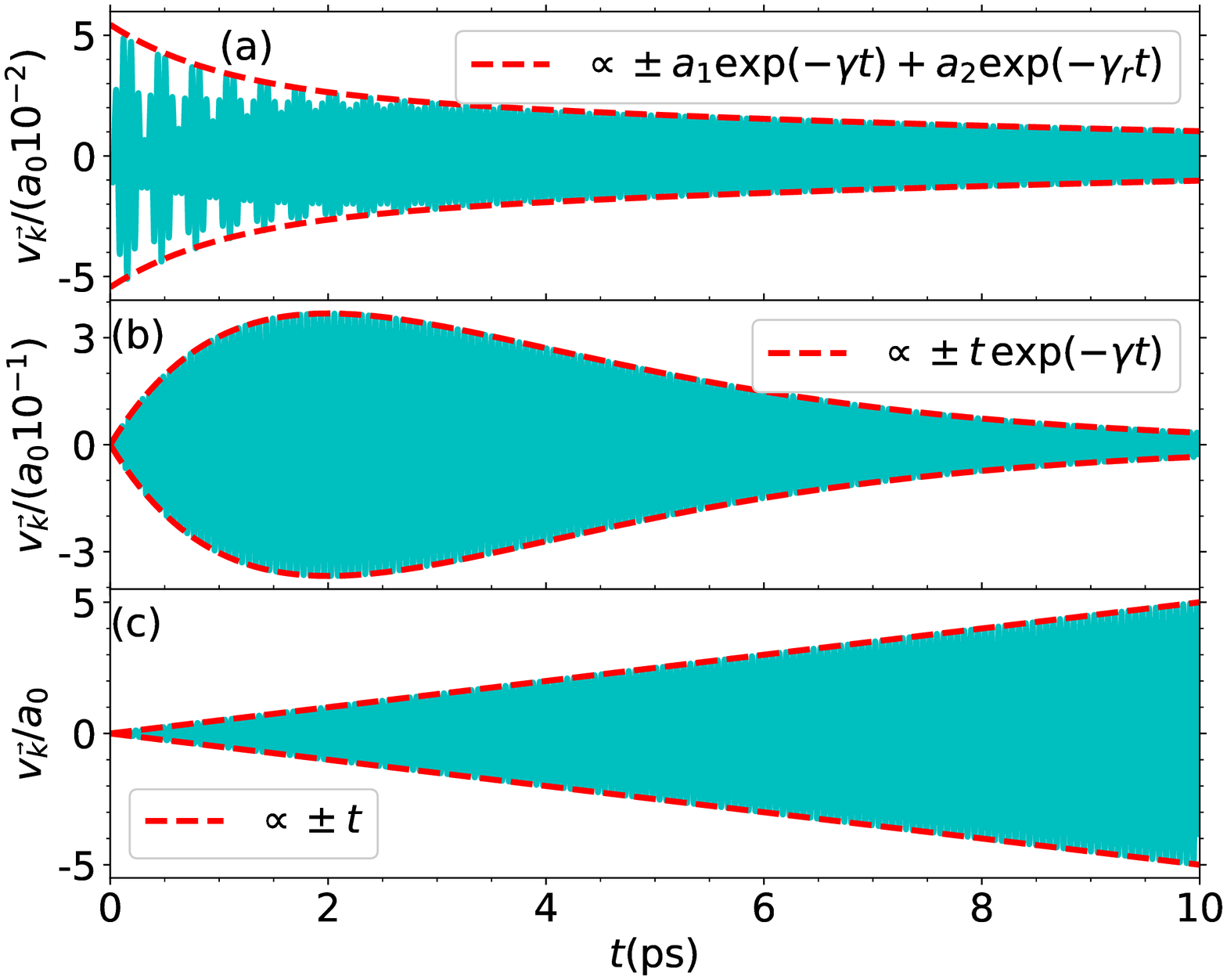}
  \caption{
The analytic solution $v_{\vec{k}}(t)$  according to
  Eq.~\eqref{eq:z_final_solution_gamma_r} for off-resonant driving and
        Eq.~\eqref{eq:z_final_solution_gamma_r_resonance} for resonant driving,
        respectively.
  (a) OOff-resonant driving with $\omega_0 = \SI{80}{\per\pico\second}$, $\omega_{\vec{k}} = \SI{50}{\per\pico\second}$, $\gamma =\SI{1.0}{\per\pico\second}$, and $\gamma_r = \SI{0.1}{\per\pico\second}$.
  Resonant driving for $2\omega_{\vec{k}}= \omega_0 =
        \SI{100}{\per\pico\second}$ with (b) $\gamma=\gamma_r = \SI{0.5}{\per\pico\second}$
        and (c) $\gamma=\gamma_r = 0$ (Parameters: $\kappa = \num{0.01},
        \beta_{\vec{k}} = 1$).}
  \label{fig:v_envelope}
\end{figure}

Figure~\ref{fig:v_envelope}(a) shows this general behavior for
off-resonant driving with $\omega_0 = \SI{100}{\per\pico\second}$,
$\omega_{\vec{k}} = \SI{80}{\per\pico\second}$, $\gamma =
\SI{1.0}{\per\pico\second}$, and $\gamma = \SI{0.1}{\per\pico\second}$
as cyan line. The envelope is plotted as dashed red line; it is
a the superposition of the two decays determined by
$\gamma$ and $\gamma_r$.

From Eq.~\eqref{eq:z_final_solution_gamma_r} we read off
that $v_{\vec{k}}(t)$ and $w_{\vec{k}}(t)$ depend linearly
 on the driving amplitude $a_0$ in leading order
and that $v_{\vec{k}}(t)$ and $w_{\vec{k}}(t)$
being real and imaginary part of  $\langle b_{\vec{k}}^\dagger b_{-\vec{k}}^\dagger \rangle $
are related by a phase shift of $\pi/2$.
Substituting the solution for $w_{\vec{k}}(t)$ into
Eq.~\eqref{eq:eom_u} reveals
that $u_{\vec{k}}(t) \propto a_0^2$ in lowest order.

The denominator in Eq.~\eqref{eq:z_final_solution_gamma_r} vanishes
for resonant driving $\omega_0=2\omega_{\vec{k}}$ and the
special choice $\gamma=\gamma_r$.
A closer inspection reveals that this is a removable singularity without physical significance:
Considering
the limit $\omega_0\to 2\omega_{\vec{k}}$ properly we obtain
\begin{equation}
  z_{\vec{k}}(t) =\frac{i a_0 \beta_{\vec{k}}}{2} e^{-\gamma t}
        \Big(t e^{i \omega_0 t} +\frac{1}{\omega_0}\sin(\omega_0 t)\Big) .
  \label{eq:z_final_solution_gamma_r_resonance}
\end{equation}
As known from resonant driving of harmonic oscillators,
the amplitude increases linearly as a result of the secular
term. In present of damping, the exponential decay sets in
at later times.

Figure~\ref{fig:v_envelope}(b) illustrates the envelope of the resonantly
driven system at $\gamma = \gamma_r = \SI{0.5}{\per\pico\second}$. The
envelope is given approximately by $\propto t \, \exp(-\gamma t)$
as expected from  Eq.~\eqref{eq:z_final_solution_gamma_r_resonance}.
For $\gamma=\gamma_r=0$, we indeed obtain a linearly increasing
amplitude as shown in Fig.~\ref{fig:v_envelope}(c).

On the basis of experimental findings for $\alpha$-MnTe \cite{bossiniMnTeExp2021},
we assume
that the relaxation rate $\gamma_r$ is significantly smaller than
the damping rate $\gamma$ of the driving term.
Consequently, the effect of the decay of the magnon occupation
happens on time scales much longer than the duration of
the driving pulse. This is also the justification to introduce
this decay in a phenomenological way. It is the slowest effect
and thus can be described  by a Lindbladian dynamics.

The contribution of resonantly driven
$\vec{k}$-modes  to $\delta L$ is especially large.
Hence, we want to understand their dynamics in particular.
For simplicity, we set $\gamma_r = 0$ which is justified for short
and intermediate time scales where $\gamma_r t\ll 1$.
First, we consider a driving term with constant amplitude. For this
case, we are able to perform a more comprehensive analytical calculations than
the one presented in Eq.~\eqref{eq:z_final_solution_gamma_r} by
adapting the approach in Ref.~\cite{PhysRevB.103.045132}. We
take the terms $a(t) \alpha_{\vec{k}}$ and $u_{\vec{k}}$ in the integral
in Eq.\ \eqref{eq:z_equ_simple_gammar} into account.

We focus on the slow change of $u_{\vec{k}}$ only, i.e., we
average out the fast oscillations at frequencies $\omega_0$ and $2\omega_0$.
This is done by averaging over $T_0 = \frac{2\pi}{\omega_0}$,
for details see App.\ \ref{app:constant}.
We distinguish between resonant driving and two types of off-resonant driving.
At the resonance condition $2 \omega_{\vec{k}} = \omega_0$,
the magnon occupation increases in time without limit
\begin{subequations}
\label{eq:tuned}
\begin{eqnarray}
u_{\vec{k}}(t) &=& \frac{1}{2} (\cosh(\Gamma t)-1)
\label{eq:u_res}
\end{eqnarray}
with
\begin{eqnarray}
\Gamma &:= &\frac{\beta_{\vec{k}} \omega_0}{\alpha_{\vec{k}}} J_1
\Big(\frac{2 a_0\alpha_{\vec{k}}}{\omega_0}\Big) \approx a_0 \beta_{\vec{k}}.
\label{eq:Gamma}
\end{eqnarray}
\end{subequations}
Here, $J_1(x)$ denotes the Bessel function of the first kind.
For numbers that are typically present in the experiment we can approximate it by its
linear term in the Taylor series for small argument.

For a driving frequencies slightly off-resonance
with a finite detuning
$\delta = 2 \omega - \omega_0$, we have to distinguish
between two cases depending on whether the detuning is smaller
(a) or larger (b) than the energy scale of pumping $\Gamma$
\begin{subequations}
  \begin{align}
 \intertext{(A) $\Gamma > |\delta|$}
  u_{\vec{k}}(t) = \frac{1}{2}\frac{\Gamma^2}{\Gamma'^2} (\cosh(\Gamma' t)-1)
        \label{eq:u_offresa}
        \\
  \text{with } \Gamma' := \sqrt{|\Gamma^2-\delta^2|}\\
 \intertext{(B) $\Gamma < |\delta|$}
  u_{\vec{k}}(t) = \frac{1}{2} \frac{\Gamma^2}{\Gamma'^2}(1-\cos(\Gamma' t)).
        \label{eq:u_offresb}
\end{align}
\label{eq:detuned}
\end{subequations}

\begin{figure}[tb]
  \centering
  \includegraphics[width=1.08\columnwidth]{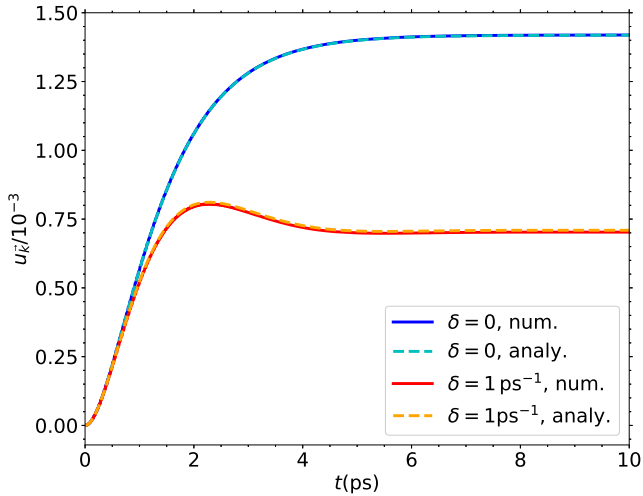}
\caption{
 The magnon occupation $u_{\vec{k}}(t)$ of a specific magnon mode
$\omega_{\vec{k}} = \SI{50}{\per\pico\second}$
for damped resonant (dark and light blue) and
damped off-resonant (red and orange) driving $\delta=\SI{1}{\per\pico\second}$.
The solid curves represent the full numerical solution while the
dashed curves represent the analytical approximations
Eqs.~\eqref{eq:tuned_damped} and \eqref{eq:detuned_damped}
with $\kappa = 0.01, \beta_{\vec k} = 1,
\Gamma= \SI{0.0753}{\per\pico\second}, \Gamma'= \SI{0.9972}{\per\pico\second}$.
}  \label{fig:specific_mode_damped}
\end{figure}

Figure~\ref{fig:specific_mode_damped} illustrates the numerical results
if the driving term is no longer constant, but damped with
$\gamma = \SI{1}{\per\pico\second}$. The
figure shows the numerically computed
dynamics of the same $\vec{k}$-mode for
resonant (blue solid curve) and off-resonant driving (red solid curve).
The numerical solution contains fast oscillations at frequency
 $2\omega_0$. Since the amplitudes are very small, these oscillations are not visible in the figure. However, Fig.~\ref{fig:specific_mode}b shows them analogously for the case $\gamma = 0$. Due to the finite phonon oscillation duration
$\approx 1/\gamma$ the magnon occupation does not grow
beyond any limit as for indefinite driving, but saturates. This behavior applies to the case of resonant and for
detuned driving.

We extend the derivation of the slowly varying contribution to
$u\kk$ to the case where the displacement oscillation
has an exponential decaying
envelope, see App.\ \ref{app:damped}. This case, however, is more
subtle than the previous case so that
an analytical derivation is only possible in leading order of $a_0$, so that
we henceforth use $\Gamma=a_0\beta\kk$, cf.\ Eq.\ \eqref{eq:Gamma}.
This assumption allows us to derive a differential equation for
the slowly varying part of $u\kk$, see App.\ \ref{app:damped}.
But, unfortunately no closed analytical expression for its solution
could be found. If we assume that the values of $u\kk$ are such
that they fulfill $2u\kk\ll 1$ we can establish
\begin{equation}
u\kk(t) = \frac{\Gamma^2}{4\gamma^2}(1-\exp(-\gamma t))^2 \underset{t\to\infty}{\to}
\frac{\Gamma^2}{4\gamma^2}
\label{eq:tuned_damped}
\end{equation}
for the resonant case and for the detuned case
\begin{subequations}
\begin{align}
u\kk(t) &= \frac{\Gamma^2}{4(\gamma^2+\delta^2)}|\exp(i\delta t)-\exp(-\gamma t)|^2
\\& \underset{t\to\infty}{\to} \frac{\Gamma^2}{4(\gamma^2+\delta^2)}
\end{align}
\label{eq:detuned_damped}
\end{subequations}
for small detuning $|\delta|\ll\omega_0$. Both show that $u\kk$ saturates to a
finite value
because we do not consider magnon relaxation $\gamma_r$ in this analysis.
To neglect $\gamma_r$ in the analysis for short time scales is justified
because its effect generically sets in on longer time scales.
Note that the saturation value is not necessarily reached monotonically in the
case of finite detuning, see Fig.\ \ref{fig:specific_mode_damped}(a).
It is interesting that slow oscillations with the frequency given by
the detuning occur.

\subsection{Pulse induced changes of sublattice magnetization}
\label{sec:pulse-magnetization}

The effect of the driving laser pulse is characterized by three parameters
in Eq.\ \eqref{eq:DJ3-vs-t}.
The parameter $a_0$ determines the oscillation strength induced by the
coherent optical phonon
at frequency $\omega_0$ which has been excited by the laser. It
is damped by the relaxation rate $\gamma$. Here we do not yet consider the
long-time decay due to the relaxation rate $\gamma_r$.
In the next three sections, We analyze the effect of each of these
parameters on the changes
of the sublattice magnetization $\delta L(t)$ introduced in
Eq.\ \eqref{eq:delta-L-def}.

\begin{figure}[htb]
  \includegraphics[width=1.08\columnwidth]{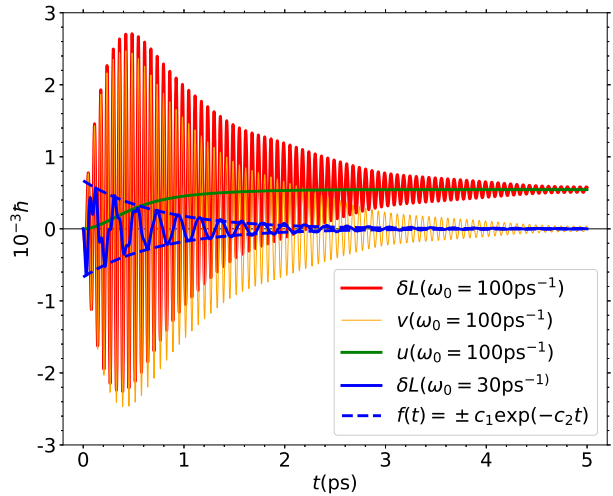}
  \caption{The change of the sublattice magnetization oscillates in time depending
        on the parameters of driving: the strength of driving $a_0$, the damping $\gamma$,
        and the
        driving frequency $\omega_0$. The curves are computed for $\kappa = 0.01,
        \gamma=\SI{1.0}{\per\pico\second}$.
        An envelope of the amplitude can be fitted
        $f(t) = \pm c_1 \exp(-c_2 t)$ with $c_1 = \num{6.7(1)e-4}, c_2
        = \SI{1.34(5)}{\per\pico\second}$. Exemplarily, the two main contributions of $\delta L$ are shown for $\omega_0=\SI{100}{\per\pico\second}$ (red curve),
              see Eqs.~\eqref{eq:delta-L-def} and \eqref{eq:uvw_k}.
              The slowly evolving contribution (green curve) consists of the magnon occupation
      $u_{\vec{k}}$
              summed over all $\vec{k}$-modes in Eq.~\eqref{def:ut}. The oscillating part consists of
              the summed contributions $v_{\vec{k}}$ in Eq.~\eqref{def:vt} (orange curve) (Parameter: $M=1000, \gamma_r = 0$).}
  \label{fig:temporal_evolution_delta_L}
\end{figure}

In Fig.~\ref{fig:temporal_evolution_delta_L}, two generic curves are depicted
for the same $\gamma$ and amplitude $a_0$, but different driving frequencies
$\omega_0$.
No relaxation $\gamma_r$ has been considered here.
Coherent oscillations occur at the driving frequency.  The envelope function
consists of fast building while the driving field is active and a consecutive slow
decay.
For the faster driving frequency $\omega_0$  a build-up phase and
 a more complex decay pattern is observed. For the lower driving
frequency, we are able to fit an exponential envelope function with a
decay rate of $c_1=\SI{1.2(3)}{\per\pico\second}$ displayed
 as dashed blue line in Fig.\ \ref{fig:temporal_evolution_delta_L}.

The change of the sublattice magnetization saturates at a finite value
$\delta L (t \to\infty)$  which can be calculated by
\begin{eqnarray}
\delta L_\infty  &=& \lim_{T\to \infty} \frac{1}{T}\int_0^T \delta L (t).
\end{eqnarray}
In order to distinguish qualitatively different contributions to $\delta L$,
we plot the slowly varying contribution $u(t)$ and the oscillating one $v(t)$
defined by
\begin{subequations}
\begin{eqnarray}
\label{def:ut}
  u(t) &=& \sum_{\vec{k}} \frac{A_{\vec{k}}}{\omega_{\vec{k}}} u_{\vec{k}},
        \\
\label{def:vt}
 v(t) &=&\sum_{\vec{k}} \frac{B_{\vec{k}}}{\omega_{\vec{k}}} v_{\vec{k}}
\end{eqnarray}
\end{subequations}
in Fig.~\ref{fig:temporal_evolution_delta_L}. The saturation value
 $\delta L_\infty$ is determined by the long-time values of the
magnon occupations since the oscillating contribution of $v_{\vec{k}}$ in
Eq.~\eqref{eq:delta-L-def} averages out and
vanishes due to dephasing.

The sum of all magnon occupations $u_{\vec{k}} (t)$ is always positive and
reaches a finite value for large times after a single laser pulse.
As a result, the change of the sublattice magnetization
approaches a finite value as well. Since we have neglected interactions
between different $\vec{k}$-modes the constant saturation cannot decay.
In experiments, however, slow relaxation processes induce a slow decay of
$\delta L_\infty\to 0$ requiring to include an additional phenomenological
relaxation rate $\gamma_r$, see below.

\subsubsection{Dependence on the driving amplitude}

\begin{figure}[htb]
  \includegraphics[width=1.08\columnwidth]{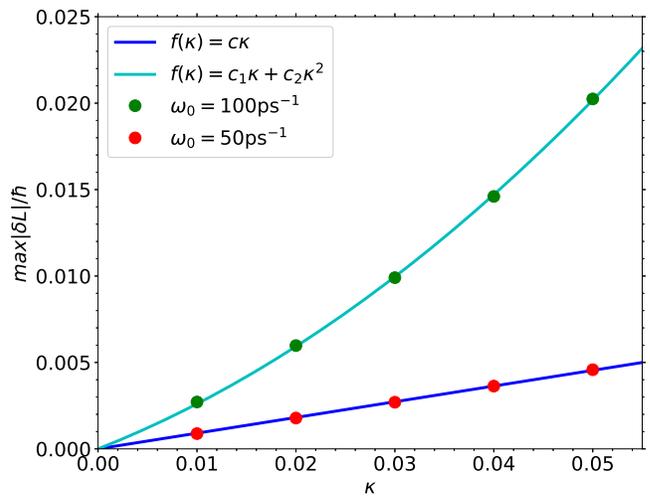}
  \caption{The maximum value of $\vert \delta L \vert$ depends linearly on the relative
        amplitude $\kappa$ of the driving, see Eq.~\eqref{eq:kappa_i},
        if the driving is small enough. Generally, non-linear contributions are also
important.
        Parameters: $\gamma = \SI{1.0}{\per\pico\second},
        M= 100, c = \num{0.0909(4)}, c_1=\num{3.6(1)}, c_2=\num{0.225(6)}$.
}
  \label{fig:M_a0}
\end{figure}

The response of the sublattice magnetization
 increases with increasing amplitude $a_0$ of the driving.
For a quantitative measure, we extracted the absolute maximum of
$|\delta L(t)|$, denoted max$|\delta L |$, and plotted it versus $\kappa$
for two driving frequencies $\omega_0$ in Fig.~\ref{fig:M_a0}. The symbols
represent the values of the numerical simulations while the solid lines
are first and second order fits, respectively.

According to Eq.\ \eqref{eq:delta-L-def}, $\delta L(t)$ consists of two
contributions: the
driving induced the magnon occupations $u_{\vec{k}}=\langle n\kk\rangle$ and the
major oscillatory part given by the sum over all $v_{\vec{k}}=\Re \langle b^\dag\kk
b\kk \rangle$.
While the contribution stemming from all $u_{\vec{k}}$ is small and mainly determines
$\delta L_\infty$, $|\delta L(t)|$ is strongly influenced by $v_{\vec{k}}$.

In order to study the dependence on $a_0$ in leading order, we inspect the
differential equations
\eqref{eq:eom_u}-\eqref{eq:eom_w}. For very small magnon occupations,
$u_{\vec{k}}\ll 1$,
one finds  $v_{\vec{k}},w_{\vec{k}}\propto a_0$ from Eq.~\eqref{eq:w_k} and,
consequently, $u_{\vec{k}}\propto a_0^2$ through Eq.~\eqref{eq:u_k}.
Therefore, we expect a parabolic fit $c_1 a_0 + c_2 a_0^2$ describing
the $a_0$ dependencyeof max$|\delta L |$ very accurately, as demonstated by
the solid lines in Fig.~\ref{fig:M_a0}. We can fit the max$|\delta L |$
dependence on $a_0$ with $c a_0$
for $\omega_0 = \SI{50}{\per\pico\second},
\gamma = \SI{1.0}{\per\pico\second}$, with $c=\num{0.0909(4)}$,
and added the fit curve as solid blue line in Fig.~\ref{fig:M_a0}.
For $\omega_0 = \SI{100}{\pico\second}$, a parabolic fit $c_1 a_0 + c_2 a_0^2$ is
needed
with $c_1=\num{3.5(1)}, c_2=\num{0.225(6)}$; it is added as cyan solid lines
confirming our analysis of Eq.~\eqref{eq:eom_uvw}.

\subsubsection{Dependence on the driving duration}

\begin{figure}[tb]
  \includegraphics[width=\columnwidth,trim=30 0 50 0,clip]{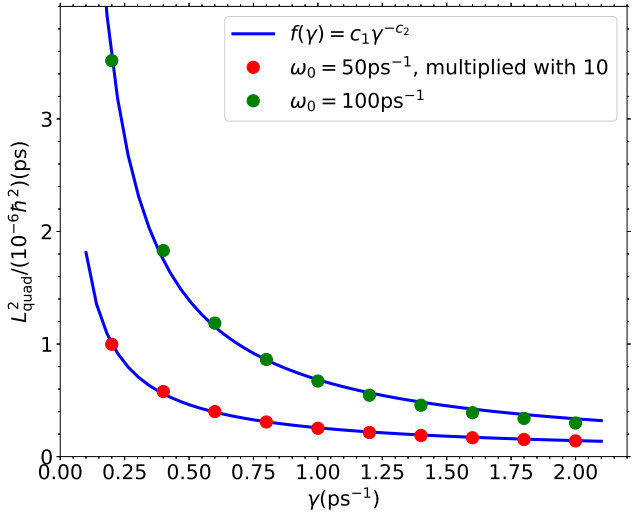}
  \caption{The value $L^2_\text{quad}$ decreases for increasing decay rate $\gamma$ of
the
        driving term roughly like $1/\gamma$.
        The data for $\omega_0=\SI{50}{\per \pico\second}$
        is multiplied by 10 for better visibility. The fit parameters
        are  $c_2 = 0.850\pm 0.009 $ for $\omega_0=\SI{50}{\per \pico\second}$
        and $c_2 = 1.02\pm 0.01$ for  $\omega_0=\SI{100}{\per \pico\second}$
        (Parameter: $M=100, \kappa=\num{0.005}$.)
  }
  \label{fig:M_gamma}
\end{figure}

The driving duration  $\propto 1/\gamma$ is parameterized
by the decay rate $\gamma$  of the coherent phonon
mode responsible for the periodic modulation of $\delta J_3(t)$
as stated in Eq.~\eqref{eq:DJ3-vs-t}.
In order to quantify the total effect of the fluctuating part of the
sublattice magnetization relative to the steady state we define
\begin{equation}
  L^2_\text{quad} := \int_0^\infty (\delta L(t)-\delta L_{\infty})^2 dt.
  \label{eq:L_quad}
\end{equation}
Note that we deduct the saturation value $\delta L_{\infty}$ in order to
ensure convergence of the integral in spite of its infinite upper limit.
So the quantity $L^2_\text{quad}$ measures in particular the oscillatory
part of $\delta L$.

In Fig.~\ref{fig:M_gamma}, the data points for $L_\text{quad}$
are depicted as colored dots as function of $\gamma$
for two frequencies $\omega_0$ and a fixed driving amplitude $a_0$ .
The solid lines are fits of the form
\begin{equation}
  f(\gamma) = c_1 \gamma ^{-c_2}
\end{equation}
with an exponent $c_2\approx 1$. The simulated data
and the fits with exponents of about $1$ agree quite well. Since the duration
of the displacement oscillation is $\propto 1/\gamma$ this result reflects the fact that the
response of the system grows proportional to the time the driving is applied.
This makes sense since the amount of energy which can be deposited by the
oscillationg phonon grows linearly with the time it lasts.

\begin{figure}[tb]
  \includegraphics[width=1.08\columnwidth]{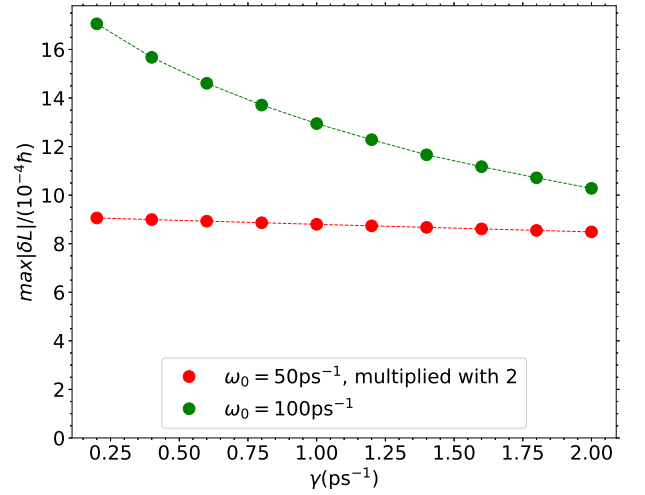}
  \caption{$\text{max}|\delta L|$ vs the damping rate $\gamma$ for two values of $\omega_0$.
  The data for $\omega_0=\SI{50}{\per\pico\second}$ is multiplied by 2 for better visibility.
  (Parameter: $M=100, \kappa=\num{0.005}$).
  }

  \label{fig:M_gamma_Lmax}
\end{figure}

Since we subtract the long-time saturation value $\delta L_{\infty}$
 governed by the sum over magnon occupations $u_{\vec{k}}$ in Eq.\eqref{eq:L_quad},
$L^2_\text{quad}$ is mainly sensitive to the two-magnon off-diagonal expectation value
$v_{\vec{k}}$.  Some analytic insight into the dynamics of the full problem can be
gained by our approximate solution
for $v_{\vec{k}}(t)$ stated in Eq.~\eqref{eq:z_final_solution_gamma_r}.
Since the time-dependent parts in Eq.~\eqref{eq:z_final_solution_gamma_r}
only depend on the magnon energy $\omega_{\vec{k}}$ one can define an
auxiliary effective DOS $\tilde\rho(\omega)$,
\begin{eqnarray*}
\tilde\rho(\omega) &=& \frac{1}{N \omega}\sum_{\vec{k}}
\beta_{\vec{k}} B_{\vec{k}}  \delta(\omega-\omega_{\vec{k}})
\end{eqnarray*}
to convert the $k$-summation in Eq.\ \eqref{eq:delta-L-def}
for the $v_{\vec{k}}(t)$ contribution into a
integration over frequency.
The resonantly driven magnons contribute the most
to the integral: Since the amplitude is proportional to $1/\gamma$ and the
width in frequency of the modes with the highest
amplitude is proportional to $\gamma$, the overall $\gamma$ dependency cancels out in
$(\delta L(t)-\delta L_{\infty})$
This is clearly shown in Fig.~\ref{fig:M_gamma_Lmax} where
one can discern that the maximum of $\delta L$ depends only very weakly on $\gamma$
especially for smaller $\omega_0$.
Then we are left with exponential decay
of $v_{\vec{k}}(t)$
with rate $\gamma$ implying
that the magnetic response lasts for about $1/\gamma$, see Sect.~\ref{sss:dephasing}. Due to the integration in Eq.~\eqref{eq:L_quad} this implies $L^2_\text{quad} \propto 1/\gamma$ as demonstrated in Fig.~\ref{fig:M_gamma}.

\subsubsection{Dependence on the driving frequency $\omega_0$}

\begin{figure}[tb]
  \includegraphics[width=1.0\columnwidth]{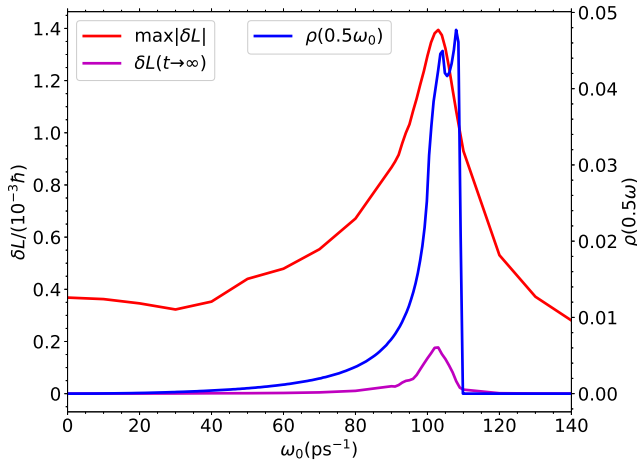}
  \caption{The maximum value of $|\delta L|$ and the saturation value
        $\delta L(t\to\infty)$ show a behavior similar
        to the magnon DOS except for a factor 2 in the frequency
        because of the resonance condition $2\omega\kk \approx \omega_0$.
        ($M=100, \kappa = 0.005, \gamma=\SI{1}{\per \pico\second}$).}
  \label{fig:M_omega0}
\end{figure}

Fig.~\ref{fig:temporal_evolution_delta_L} already shows that
the change of the sublattice magnetization strongly depends on
the driving frequency $\omega_0$ for the same driving amplitude $a_0$ and
driving decay $\gamma$.
For fixed $a_0$ and $\gamma$ we scan the driving frequency $\omega_0$ and
extract ${\rm max}|\delta L|$ as well as the saturation value $\delta L_\infty$.
The generic results of this scan are depicted in Fig.~\ref{fig:M_omega0}.

Even though the sublattice magnetization and all related quantities
are sums over all $\vec{k}$-modes it is obvious that the modes
in resonance or very close to it contribute that most to $\delta L$.
Hence, one expects that the response of $\delta L$
follows roughly the DOS of the magnons because a high
density of modes with $2\omega\kk\approx\omega_0$ is favorable
for a strong effect in $\delta L$. As this argument implies
there is a factor of 2 between the DOS and the behavior of $\delta L$
because the resonance condition holds for pairs of magnons.
Indeed,  Fig.~\ref{fig:M_omega0} clearly illustrates that these
ideas are correct, at least on a qualitative level.
As expected, ${\rm max}|\delta L|$ and $\delta L_\infty$ peak at the
van-Hove singularities of the magnon DOS.

Interestingly, the saturation value ${\rm max}|\delta L|$ becomes
significant only in the vicinity of the peaks of the van-Hove
singularities while ${\rm max}|\delta L|$ acquires significant
values also away from the prominent peaks of the DOS.
We attribute this to the fact that saturation value is large only
at or close to resonance, see Eq.\ \eqref{eq:detuned_damped}.
A mode is far from resonance if $|\delta| \gg \gamma$ which is
the case for most modes in view of the typical small values of $\gamma$.
In contrast, ${\rm max}|\delta L|$ depends on the oscillatory
contributions of all modes at small or moderate times, see Eq.\
\eqref{eq:z_final_solution_gamma_r}. Hence it does not depend
so strongly on the DOS of magnons.

\subsubsection{Long-time decay of the change of the sublattice magnetization}
\label{sss:dephasing}

We still consider the physical situation without phenomenological
magnon decay, i.e., for $\gamma_r=0$. Still, the change of the sublattice magnetization
does decay in seeming contrast to the dynamics of single modes.
The decay of $\delta L$ in time results from dephasing, i.e., from
the fact that all $\vec k$-modes contribute but they display oscillations
with a broad range of frequencies which quickly become out-of-phase so that
the signal decreases. The time scale on which this dephasing takes place
is roughly estimated by $\tau_\text{dephas} \approx 1/(2\omega_\text{max})$
where $\omega_\text{max}$  is the maximum frequency of the magnon dispersion and the
factor
2 stems from the fact that pairs of magnons are created. In generic experimental
set ups, $\tau_\text{dephas}$ is much smaller than the displacement oscillation duration $1/\gamma$
so that the signal dies out extremely fast once the
driving oscillation disappeared.
In turn, this implies that the characteristic time scale on which
$\delta L$ vanishes is expected to be proportional to the duration of the
driving term.

\begin{figure}[tb]
  \includegraphics[width=1.1\columnwidth]{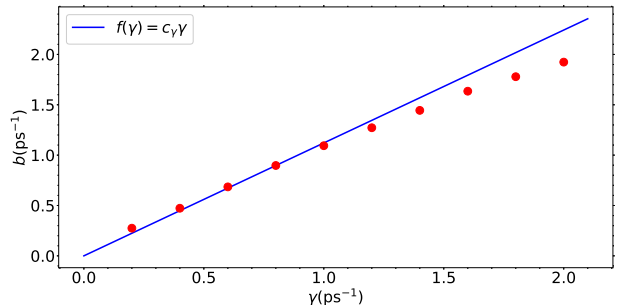}
  \caption{Effective dephasing rate $b$ of the envelope of oscillatory part
        of the sublattice magnetization as function of the decay rate of the driving force.
        For small $\gamma$ the relation is linear as expected in the limit where the
        actual dephasing rate is very large, see main text. The prefactor
        is of the order of unity $c_\gamma = \num{1.12(2)}$.
        ($M=100, \kappa=0.005,\omega_0=\SI{30}{\per\pico\second}$)}
  \label{fig:decay_gamma_b}
\end{figure}

We want to put this hypothesis to test. As shown in
Fig.~\ref{fig:temporal_evolution_delta_L},
$\delta L(t)$ approaches a finite saturation value $\delta L_\infty$ for $\gamma_r =0$
while the oscillatory part decays in time. We separate the oscillatory part
of $\delta L$  and fit its envelope by an exponential
\begin{equation}
 e(t) = ce^{-b(\gamma)t}
\end{equation}
from which we read off the effective
dephasing rate $b(\gamma)$. The resulting values are displayed
in Fig.~\ref{fig:decay_gamma_b} as function of $\gamma$. We stress that
this analysis was done for a relatively small driving frequency
$\omega_0 = \SI{30}{\per\pico\second}$ which
is far away from the van-Hove singularites where the two peaks in the DOS
 introduce an additional time dependence as shown by the red curve in
Fig.~\ref{fig:temporal_evolution_delta_L}. Then a mono-exponential analyis
of the characteristic dephasing rate is not possible.

For a frequency in the featureless range of the magnon DOS we find
our hypothesis supported in Fig.~\ref{fig:decay_gamma_b} displaying
a linear dependence $b\propto\gamma$. This means that the time scale
of the decay of the oscillations in $\delta L$ is dominated by
the time scale of the
driving oscillation duration. Note that the proportionality
factor $c_\gamma$ is about unity so that both time scales coincide
the regime of small $\gamma$. Only for larger values of
$\gamma$ some downward curvature appears. This feature is not unexpected
either because for $\gamma\to\infty$ one expects that the limiting
process of signal decay is the dephasing on the time scale
$1/(2\omega_\text{max})$ so that $b(\gamma)$ saturates at some value
$b_\infty = \lim_{\gamma\to\infty}(b(\gamma))$ of the order of
$2\omega_\text{max}$.

\subsubsection{Time evolution of the total energy}

A driven system acquires energy through the driving. So it is an important question
how much energy is pumped into it and on which parameters this effect depends.
We define the total energy  energy $E$ per spin by
\begin{eqnarray}
\label{eq:E_def}
E(t)&=& \frac{1}{N} \left(\langle H(t) \rangle - J_1S(E_d + \Delta E)\right)
\end{eqnarray}
where $H(t)$ is given by  Eq.~\eqref{eq:energy_timedependent}.
In the above definition we subtracted the trivial
energy offset of $H_0$. While $E(t)$ initially oscillates and
 increases on average it reaches a saturated value
for times $t\gg 1/\gamma$ which represents  its long-time limit.

\begin{figure}[tb]
  \includegraphics[width=1.08\columnwidth]{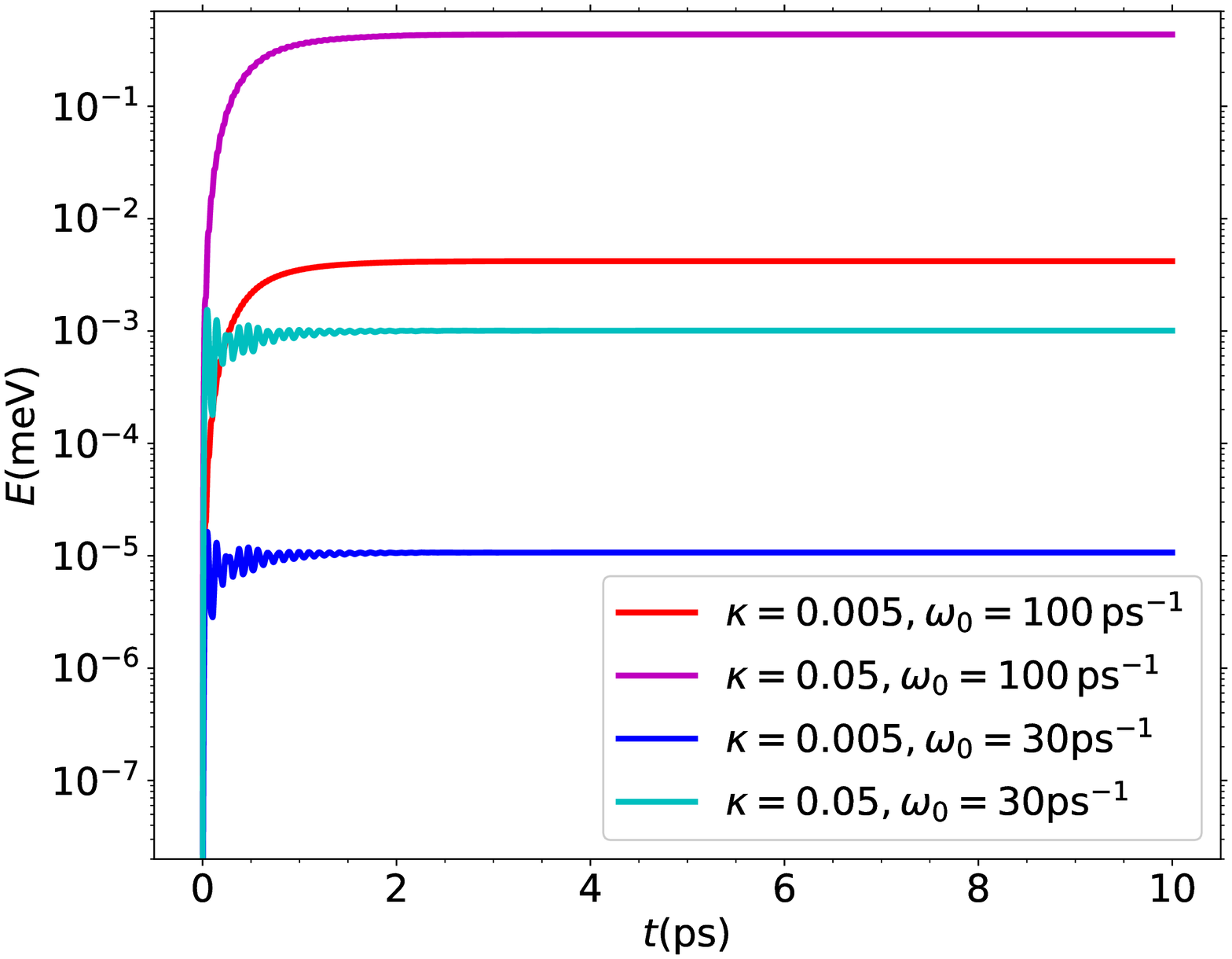}
  \caption{
  The energy per site $E(t)$ for different combinations of $\kappa$ and $\omega_0$. The energy
reaches
        a saturation value for $t\to\infty$ since the driving oscillation is damped with rate
        $\gamma = \SI{1}{\per\pico\second}$, i.e., the oscillation is effectively of finite
duration.
        All curves start at zero energy since we start from the system at zero temperature.
         (Parameter: $M=100, \gamma_r=0$
         ).
        }
  \label{fig:energy}
\end{figure}

Figure \ref{fig:energy} depicts the time evolution of $E(t)$ for
two different driving frequencies and two different
driving amplitudes in Fig.~\ref{fig:energy}.
The saturation value of the energy depends on the driving frequency and the
amplitude of the driving. If the resonance condition
is fulfilled or almost fulfilled for many of $\vec{k}$-modes,
quantified by the magnon DOS in Fig.\ \ref{fig:dispersion_DOS}(b),
the energy uptake is facilitated and higher values of energy are reached.
Thus, even a small driving with $\kappa=0.005$ leads to a higher excitation
if it is done at frequency at the peaks of the DOS
than the ten times larger driving at small values of the DOS.

For the stronger driving one can discern oscillations for short times. They
are also present for the weaker driving, but much smaller. We presume that
the driving at around the DOS maxima leads to substantial contributions from
many magnons so that the signal is better averaged and hence does not fluctuate
so strongly.
The amplitude of the oscillation is much smaller than the increasing slowly varying component of $E(t)$
and, therefore, not visible in the figure.

\begin{figure}[t]
  \includegraphics[width=1.08\columnwidth]{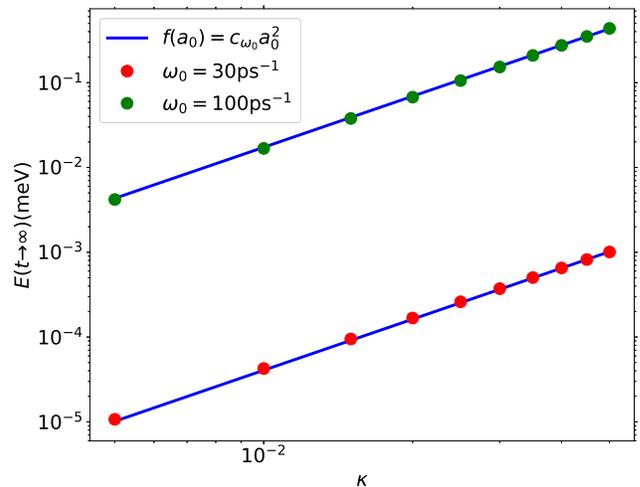}
  \caption{
  The saturation value of the energy per site $E(t\to\infty)$  depends
        quadratically on the relative amplitude $\kappa$ of the driving. The fit parameters
        are $c_{\omega_0} = \SI{ 0.406+-  0.001}{\per\pico\second}$ for
        $\omega_0=\SI{30}{\per\pico\second}$ and
        $c_{\omega_0} = \SI{173.0 +- 0.5}{\per\pico\second}$
        for $\omega_0=\SI{100}{\per\pico\second}$ (Parameter: $M=100, \gamma=\SI{1.0}{\per\pico\second}$).
        }
  \label{fig:E_a0}
\end{figure}

We already demonstrated in Fig.\ \ref{fig:M_omega0}
that the change of the sublattice magnetisation reaches
a maximum for $\omega_0=\SI{100}{\per\pico\second}$.
Additionally, we selected $\omega_0 =  \SI{30}{\per\pico\second}$.
For both frequencies, we investigate the dependence of
the saturation value $E(t\to\infty)$ on the driving amplitude $a_0$.
The data shown in Fig.\ \ref{fig:E_a0} is perfectly described by
$E(t\to\infty)\propto a_0^2$.
This could be naively expected in
analogy to a driven classically harmonic oscillator where
the energy absorption is proportional to the square of the
driving amplitude or, alternatively, as direct
implication of Fermi's Golden Rule.

Inspecting the details of our calculation, we observed
already earlier that $u\kk$ grows quadratically with $a_0$, see the discussion
after Eq.\ \eqref{eq:z_final_solution_gamma_r}.
It is particularly obvious in the explicit expressions for the
saturated limits of $u\kk$ in Eqs.\ \eqref{eq:tuned_damped}
and \eqref{eq:detuned_damped}, we recall $\Gamma\propto a_0$.
In the long-time limit, the time dependent contribution $X(t)$ to the Hamiltonian
vanishes  exponentially $\propto \exp(-\gamma t)$
so that only $H_0$ contributes to $E(t)$ for $t\to\infty$.
 Consequently, only $u_{\vec{k}}$ enters in the expectation value according to
Eq.~\eqref{eq:H0-diagonal}
and hence this value depends quadratically on the driving amplitude.

\section{Making contact to experiments}
\label{sec:contact-to-experiments}

To connect our calculations closer to experiments,
we include a finite relaxation rate $\gamma_r$ that parametrizes
relaxation processes beyond the Hamiltonian \eqref{eq:hamiltonian}
in an effective Lindblad equation. The reservoir consists of all the
phonons in MnTe.

In addition, we investigate whether decisive qualitative differences occur
if the optical phonon couples to the other exchange couplings in the system,
namely $J_1$ or $J_2$.

\subsection{Effect of the magnon relaxation}
\label{sec:with_relaxation}

The linear spin wave theory employed so far neither contains neither any scattering between
the magnon modes nor does it contain relaxation terms which allow the system
to reach its initial equilibrium again after a long time. In order to provide
a theoretical description that includes
relaxation back to the initial equilibrium state prior to the photo-excitation,
 we have already introduced an additional relaxation rate $\gamma_r$
in Eqs.~\eqref{eq:eom_uvw}. In this section, we consider a finite value of
$\gamma_r>0$
explicitly. The derivation of the  approximate {analytic solution is lengthy, but can
be found in App.\ \ref{app:relax}.

\begin{figure}[htb]
  \includegraphics[width=1.08\columnwidth]{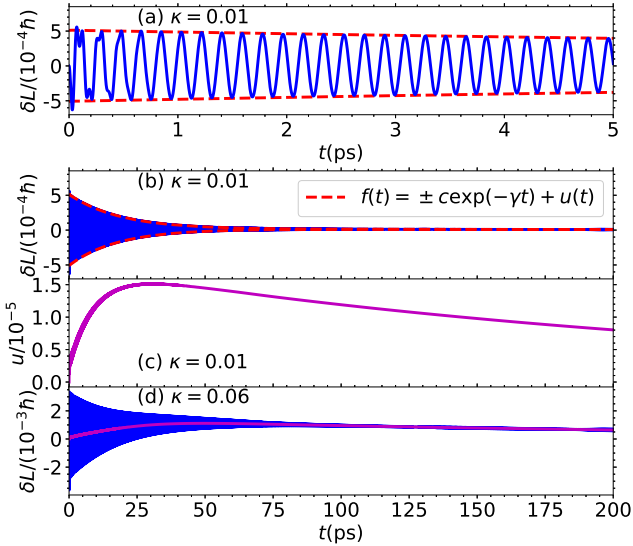}
  \caption{Evolution of $\delta L(t)$ including relaxation parametrized by
        $        \gamma_r = \SI{3.9e-3}{\per\pico\second}$. The other parameters are also
        adapted to experiment \cite{bossiniMnTeExp2021}:
        $\omega_0=\SI{33.6}{\per\pico\second}, \gamma =\SI{0.055}{\per\pico\second}, M=2200$.
        Panel (a) focuses on the short-time dynamics for $\kappa = 0.01$ to illustrate the
        fast oscillations. Panel (b) shows the long-time dynamics. Panel (c) shows the
        contribution of the slowly varying contribution $u(t)$ of $\delta L(t)$ of panel (b)
        which decays like $\exp(-\gamma_r t)$; note the difference of scales of the $y$ axes.
        Panel (d) also displays $\delta L(t)$ and the slow contribution $u(t)$, but for
larger
         relative driving amplitude $\kappa = 0.06$.}
  \label{fig:delta_L_gammar}
\end{figure}

Figure~\ref{fig:delta_L_gammar}
shows $\delta L(t)$ and $u(t)$ for
typical experimental driving frequencies and coherent lattice oscillation
durations in $\alpha$-MnTe
\cite{bossiniMnTeExp2021}. Since the driving amplitude $a_0$ is unknown, we perform
calculations for $\kappa = 0.01$ and $\kappa = 0.06$ to illustrate the possible
outcomes
for smaller and for larger driving amplitude.

Figure~\ref{fig:delta_L_gammar} shows $\delta L(t)$ as blue curves. The oscillations
of $\delta L(t)$ are so fast that they can only be resolved on time scale of panel (a).
The contribution $u(t)$ to $\delta L(t)$ is depicted separately in
Fig.~\ref{fig:delta_L_gammar}(b)
 to illustrate the relaxation due to $\gamma_r$ visible on the long-time range. For
$\kappa=0.01$,
 the contribution of the magnon occupation is very small compared to the oscillatory
contribution $v(t)$.
Therefore, including an addtional magnon relaxation term has only a minor effect on
$\delta L(t)$.  Since $\gamma = \SI{0.055}{\per\pico\second} >\gamma_r=
\SI{3.9e-3}{\per\pico\second}$, we observe
an initial increase of the amplitude due to the driving of the system up to the
time scale $T_d\approx 1/\gamma$
before the decay sets in. Then $\delta L(t)$ is determined only by
$u(t)$
as discussed above. At the larger time scale, $t>1/\gamma_r$,
 the magnon occupation $u(t)$ exponentially decays proportional to $\exp(-\gamma_r t)$
so that $\delta L(t)$ vanishes asymptotically for $t\to \infty$.

Figure~\ref{fig:delta_L_gammar}(c) depicts
$\delta L(t)$ for a larger driving amplitude  $\kappa = 0.06$.
In contrast to the results for $\kappa=0.01$, $u(t)$ shown as violet curve becomes
significantly larger relative to the oscillatory component. This is consistent with the linear
dependence of
the oscillatory component on the amplitude
$a_0$, while the slowly varying component grows quadratically in $a_0$.

For the realistic parameter regime of $\gamma>\gamma_r$, most
of the previous analysis of the case at $\gamma_r = 0$
continues to apply in the presence of $\gamma_r$.
Especially, $\mathrm{max}|\delta L|$ does not change significantly
if $\gamma \gg \gamma_r$ since it is dominated by the short-time dynamics
hardly affected by a small value of $\gamma_r$.
Therefore, we do not investigate the effect of $a_0$ and $\gamma$
on $\delta L$ for finite $\gamma_r$ again.
Only the decreasing envelope of the amplitude of $\delta L$
becomes more complicated, but shows the expected behavior:
the mono-exponential decay is replaced by a bi-exponential decay
with two decay rates $\gamma$ and $\gamma_r$. The latter
determines the behavior for long times.
We stress that including $\gamma_r$, implies that
the dynamic system approaches its fixed point  given by
$u_{\vec{k}}(t) = v_{\vec{k}}(t) = w_{\vec{k}}(t) = 0$.
The magnetic energy pumped into the Heisenberg model by the lattice
driving is eventually dissipated into the rest of the system
governed by the relaxation rate $\gamma_r$.

\subsection{Dynamics from the modulation of other exchange couplings}
\label{sec:with_other_deltaJ}

So far, we focused on the effect of a modulated coupling strength $J_3(t)$.
In general, a coupling of the optical phonon to the exchange couplings
$J_1$ or $J_2$ is also possible \cite{bossiniMnTeExp2021}.
The question arises whether the modulation
of these couplings is qualitatively different compared to the modulation of
$J_3$ studied so far.

For this purpose, only slight modifications of the theory
are necessary. The differential equations~\eqref{eq:eom_uvw} are
unchanged, but the prefactors $\alpha_{\vec{k}}$ and $\beta_{\vec{k}}$
have to be modified according to
\begin{subequations}
 \begin{align}
   \alpha_{\vec{k}} &= \frac{A_{\vec{k}}}{\omega_{\vec{k}}}
        \left( 2 - 2\frac{B_{\vec{k}}}{A_{\vec{k}}} \cos(k_c) \right)
        \\
   \beta_{\vec{k}} &= \frac{A_{\vec{k}}}{\omega_{\vec{k}}}
        \left( -2 \frac{B_{\vec{k}}}{A_{\vec{k}}} + 2 \cos(k_c) \right)
    \end{align}
\end{subequations}
for discussing a driving via $\delta J_1(t)$ and to
\begin{subequations}
 \begin{align}
   \alpha_{\vec{k}} &= \frac{A_{\vec{k}}}{\omega_{\vec{k}}}
        \left( 6 - 2\frac{B_{\vec{k}}}{A_{\vec{k}}} \gamma_\Delta(\vec{k}) \right)
        \\
   \beta_{\vec{k}} &= \frac{A_{\vec{k}}}{\omega_{\vec{k}}}
        \left( -6 \frac{B_{\vec{k}}}{A_{\vec{k}}} + 2 \gamma_\Delta(\vec{k}) \right).
 \end{align}
\end{subequations}
when considering a driving via $\delta J_2(t)$.

\begin{figure}[htb]
  \includegraphics[width=1.08\columnwidth]{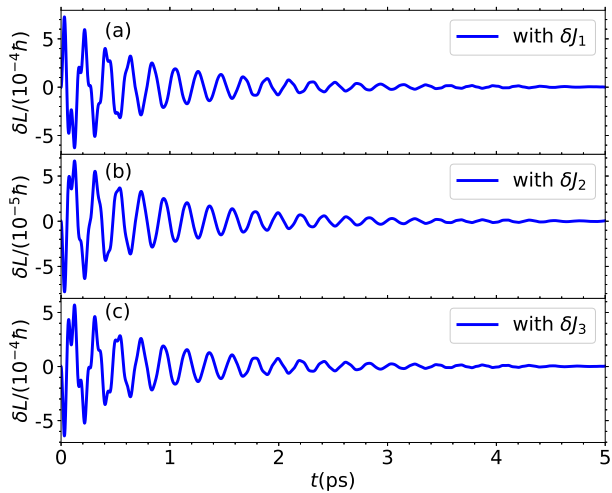}
  \caption{Evolution of $\delta L$ for modulated couplings
        (a) $J_1$, (b) $J_2$, and (c) $J_3$ with relative amplitudes $\kappa_i$.
        Parameters: $\kappa_i = \num{0.01}, M = \num{200},
        \omega_0 = \SI{30}{\per\pico\second},\gamma = \SI{1.0}{\per\pico\second},\gamma_r=0$.}
  \label{fig:diff_deltaJ}
\end{figure}

Figure~\ref{fig:diff_deltaJ} illustrates the change of the sublattice magnetization
$\delta L$ for all three cases; note the different scales of the $y$ axes.
In these calculations, we set $\omega_0 =\SI{30}{\per\pico\second}$ and
$\gamma = \SI{1}{\per\pico\second}$ and neglect the  magnon relaxation $\gamma_r$.
The data of Fig.~\ref{fig:diff_deltaJ}(c) is the same as in
Fig.~\ref{fig:temporal_evolution_delta_L}.
All curves in Fig.~\ref{fig:diff_deltaJ} are very similar even though the coupling
of the optical phonon is mediated  by different Heisenberg terms.
This is a consequence of the identical analytic structure of the
differential equations governing the dynamics of the system.

However, the amplitude of $\delta L$ is determined by the maximum relative deviation
$\mathrm{max}|\delta J_i|$ and  the product
$B_{\vec{k}}\beta_{\vec{k}}$
since $v_{\vec{k}}$ is proportional to $\beta_{\vec{k}}$ (Eq.~\eqref{eq:z_final_solution_gamma_r}) and is multiplied with $B_{\vec{k}}$ to calculate $\delta L$ (Eq.~\eqref{eq:delta-L-def}). Because $\beta_{\vec{k}}$
contain the number of nearest neighbors,
a smaller value of $J_i$ can be compensated by a large number of nearest neighbors.
Therefore, with the parameter set used in this work, the strength of driving is
approximately the same for $J_1$ and $J_3$ using the same relative coupling amplitude
$\kappa_1=\kappa_3 = \num{0.01}$. In contrast, the effect of a modulated
coupling $J_2$ is significantly smaller.

In addition, the change of the sublattice magnetization
induced by $\delta J_1$ has the opposite sign compared to
the change caused by $\delta J_2$  or $\delta J_3$.
This effect is traced back to the sign of the product
$\beta_{\vec{k}} B_{\vec{k}}$ in Eq.\ \eqref{def:vt}
which is positive for $J_1$ and negative for $J_2$ and $J_3$.

In summary, the change of the sublattice magnetization
predicted by the advocated model is up to a  sign
 very similar for the modulations of all couplings $J_i$.
Hence, the analysis of the influence of the various parameters
in the driving  via $\delta J_3 (t)$  as presented in previous sections
is sufficient to establish the essential physical response of $\delta L$.


\section{Conclusions}
\label{sec:conclusion}

The research areas of magnonics and spintronics are currently attracting major interest
\cite{barma21,malki20b}. The idea to use magnetic excitations for information transport and processing is indeed very attractive because no physical object needs to be transported through the device. Hence such devices
are considered strong candidates for reducing the energy consumption
due to coherent information processing \cite{jackl17,Bertellieabd3556}.

One key issue for embedding such magnonics devices into the
established semiconductor electronics is the conversion of spin signals
into charge signals and vice versa at the highest possible operational speed. These considerations fuel  the
presently growing research fields of ultrafast magnonics and spintronics
which aim at coupling spins and charges on the femtosecond time scales minimizing as much as possible the energy dissipation. A recent trend from the experimental side involves
the optical activation of coherent phonons either via Raman scattering processes
\cite{MERLIN1997207} or via resonant pumping \cite{FoerstEtAl2011,NovaEtAl2017}. In this framework,
it is a natural choice to investigate hexagonal MnTe since in this material the
optical band-gap, i.e.,the charge degree of freedom, is naturally coupled to both
the spins \cite{PhysRevB.61.13679,Bossini_2020,hafeztorbati2020magnetic}
 and to the lattice \cite{RamanMnTe2020} in equilibrium.

Hence, in this article we pursued
the idea that light triggers coherent lattice motion which in turn
induces coherent oscillations
of the sublattice magnetization.
Aiming at $\alpha$-MnTe, we employed a Heisenberg model whose
coupling constants are determined by
data from inelastic neutron scattering \cite{PhysRevB.73.104403}.
The optically induced atomic displacements
modulate the  exchange couplings
and thereby create pairs of magnons of opposite momenta
in the considered isotropic spin model.
We studied  the dependence of the temporal evolution of the
sublattice magnetization on the variation of parameters such as
the amplitude of the modulation of the exchange coupling, the duration
of the coherent lattice oscillation,
its carrier frequency, and a phenomenological relaxation
rate of the magnons. The used parameters are chosen in the
 experimentally relevant range \cite{bossiniMnTeExp2021}.

In particular, we calculated the dynamics
 of the sublattice magnetization $L(t)$ induced by
the oscillating Heisenberg coupling $J_3$ between third-nearest neighbors.
The time-dependent deviation $\delta L (t)$ to the equilibrium value
is given by a weighted sum of the dynamics of all $\vec{k}$-modes.
Since the differential equations describing the dynamics of the
$\vec{k}$-modes remain diagonal in momentum space in linear spin wave theory
and for the assumed relaxation mechanism,
it is possible to gain analytic insight by analyzing individual $\vec{k}$-modes.
We solved the differential equations in the presence of a driving term with
some simplifying assumptions analytically as well as fully numerically.
In this way, we have explained the properties
of $\delta L (t)$ and its dependence on the external parameters.

We found that the qualitative features depend only quantitatively on
which exchange coupling is modulated by the optical
excited coherent lattice mode.
The exception to this rule is a sign change of the oscillations of $\delta L(t)$ if the coupling $J_1$ is modulated instead of $J_2$ or $J_3$.

But except for this phase shift of $\pi$ our
 observation has two implications:
(i) generally, it demonstrates the generic nature of our approach, which is
applicable to a wide class of materials
and  (ii)  in the context of $\alpha$-MnTe,
the calculated the modulation of the sublattice magnetization does
not require the precise microscopic details of
how the optically excited displacement mode influences the exchange
paths, which is presently not known.

We stress that the computed effects in the sublattice
magnetization $L(t)$ are measurable experimentally as both the longitudinal and transversal
femtosecond dynamics of $L(t)$
 can be detected by means of magneto-optical effects in a wide variety of
materials \cite{PhysRevB.89.060405,bossi17}.
It is thus important to point out that the presented calculations for $\alpha$-MnTe
can easily be adapted to any other ordered quantum antiferromagnetic system.

Furthermore, future research can extend the studied model in various directions:
(i) The relaxation mechanism of magnons as considered here does not conserve
the total spin \cite{PhysRevB.103.045132}. More elaborate Lindblad operators
can ensure the conservation of spin which holds for the dominant processes
in many systems. (ii) We assume the unperturbed state, prior to the photoexcitation,
to be in equilibrium at zero temperature. It is straightforward to include
finite temperature. (iii) In linear spin wave theory we neglected the scattering
of the magnons among themselves. Such interaction effects can be
included on the level of Boltzmann equations, see for instance
Ref.\ \cite{kalth21}.
(iv) Finally, it is conceptually interesting to study
the short coherent drive
in the vicinity of thermal and quantum phase transitions
in order to determine whether it is possible to drive
the system from one phase into the other.

The ultrafast coherent control of macroscopic magnetic states is thus
an exciting research field still in its infancy, from both the experimental
and the theoretical side as well.

\section{Acknowledgements}

We acknowledge useful discussions with Mohsen Hafez-Torbati and Bruce Normand.
This study was carried out in the International Collaborative Research Centre 160 (Project B8)
funded by the Deutsche Forschungsgemeinschaft (DFG) and the
Russian Foundation for Basic Research. Further support (GSU) by the DFG
was obtained through the projects UH 90/13-1 and UH 90/14-1. D.B. acknowledges supports from the Deutsche Forschungsgemeinschaft  (DFG) program BO5074/1-1.

\appendix

\section{Analytical approximated solution of the differential equations}
\label{app:analytic_approximation}

To calculate the dynamics analytically, we combine the terms
$v_{\vec{k}}(t)$ and $w_{\vec{k}}(t)$  in the
 complex number $z_{\vec{k}}(t) = v_{\vec{k}}(t) + i w_{\vec{k}}(t)$
obeying the differential equation
\begin{equation}
  \frac{d z_{\vec{k}}}{dt} = 2i(\omega_{\vec{k}} + a(t)\alpha_{\vec{k}}) z_{\vec{k}} -\gamma_r
z_{\vec{k}} + i f_{\vec{k}}(t)
  \label{eq:eom_z}
\end{equation}
with
\begin{subequations}
\begin{align}
f_{\vec{k}}(t) &= 2 a(t)\beta\kk (u\kk(t)+1/2)
\\
a(t)&= a_0 \cos(\omega_0 t) \exp(-\gamma t).
\end{align}
\end{subequations}
This results from Eqs.~\eqref{eq:DJ3-vs-t}, \eqref{eq:eom_v},
and \eqref{eq:eom_w}.
Neglecting $a(t)\alpha_{\vec{k}}$ relative to
$\omega_{\vec k}\gg a(t)\alpha_{\vec{k}}$
simplifies Eq.\eqref{eq:eom_z} to
\begin{equation}
  \frac{d z_{\vec{k}}}{dt} = (2i\omega_{\vec{k}} -\gamma_r) z_{\vec{k}} + i f_{\vec{k}}(t).
  \label{eq:z_solution_gammar}
\end{equation}
For the estimated parameters relevant for experiment the approximation is well justified as we verified by comparing approximate analytical with numerical results without this approximation.
The solution of Eq.~\eqref{eq:z_solution_gammar} is given by
\begin{align}
  \label{eq:z_equ_simple_gammar}
  &z_{\vec{k}}(t) =
        \\ \nonumber
        &\qquad
        i e^{(2i\omega_{\vec{k}} - \gamma_r)t} \int_0^t a(t') \beta_{\vec{k}}
        (2u_{\vec{k}} (t') + 1)         e^{-(2i\omega_{\vec{k}} -\gamma_r)t'} dt' .
\end{align}
If we assume additionally that $2u_{\vec{k}} (t') \ll 1$,
we can neglect $u_{\vec{k}} (t')$. This holds certainly for not too strong and not too long lasting driving.
Then the integral in
 Eq.~\eqref{eq:z_equ_simple_gammar} can be solved analytically
\begin{widetext}
\begin{equation}
    z_{\vec{k}}(t) = i a_0 \beta_{\vec{k}}
    \left[\frac{e^{-\gamma t}(\omega_0 \sin(\omega_0 t)
                -(2i\omega_{\vec{k}}+\gamma-\gamma_r)\cos(\omega_0 t))}
                {\omega_0^2-(2\omega_{\vec{k}} - i (\gamma-\gamma_r))^2}
    +\frac{(2i\omega_{\vec{k}} + \gamma- \gamma_r)
                e^{2i\omega_{\vec{k}} t} e^{-\gamma_r t}}
                {\omega_0^2-(2\omega_{\vec{k}} - i (\gamma-\gamma_r))^2}
    \right].
  \label{eq:z_final_solution_gamma_r-AppA}
\end{equation}
\end{widetext}

\section{Evolution of the slowly varying part of
the magnon occupation $u\kk$.}

Here we address the computation of the slow evolution
of the magnon occupation upon driving by quickly oscillating
terms. The starting point is the set of three differential
equations at each momentum $\vec{k}$ in the Brillouin zone
\begin{subequations}
\begin{align}
  \frac{d u_{\vec{k}}}{dt} &= 2 a(t) \beta_{\vec{k}} w_{\vec{k}}
         \label{eq:dgl_u} \\
  \frac{d v_{\vec{k}}}{dt} &= -2(\omega_{\vec{k}} + a(t) \alpha_{\vec{k}}) w_{\vec{k}}
         \label{eq:dgl_v} \\
  \frac{d w_{\vec{k}}}{dt} &= 2(\omega_{\vec{k}} + a(t) \alpha_{\vec{k}})
v_{\vec{k}} + f_{\vec{k}}(t)
        \label{eq:dgl_w} \\
  \text{with } \qquad f_{\vec{k}}(t)
        &:= 2 a(t) \beta_{\vec{k}} (u_{\vec{k}} +1/2)
\end{align}
\label{eq:eqns_appAuvw}
\end{subequations}
where we omitted the phenomenological magnetic relaxation $\gamma_r$.
Obviously, the momentum dependence enters only via the
dispersion $\omega_{\vec k}$. We adapt the calculation for three-fold
degenerate triplons in Ref.~\cite{PhysRevB.103.045132} to
a  system of magnons. First, we deal with a constant amplitude, i.e.,
the pulse lasts indefinitely. Second, we generalize the
result to exponentially decaying driving term $a(t)$.

\subsection{Driving by an indefinite oscillating term $a(t)$}
\label{app:constant}

We set $a(t) = a_0 \cos(\omega_0 t)$,
define $z_{\vec{k}}(t) = v_{\vec{k}}(t) + i w_{\vec{k}}(t)$, and
combine Eqs.~\eqref{eq:dgl_v}~and~\eqref{eq:dgl_w}
\begin{equation}
  \frac{d z_{\vec{k}}}{dt} = 2i (\omega_{\vec{k}} + a_0
        \cos(\omega_0 t)\alpha_{\vec{k}}) z_{\vec{k}} + i f_{\vec{k}}(t).
\end{equation}
The solution of this differential equation reads
\begin{subequations}
\begin{align}
  z_{\vec{k}}(t) &= i e^{ih_{\vec{k}}(t)} \int_0^t f_{\vec{k}}(t')
        e^{-ih_{\vec{k}}(t')} dt'
        \label{eq:solution_z}
        \\
    \text{with } h_{\vec{k}}(t) &:= 2 \int_0^t (\omega_{\vec{k}}
                + a_0 \alpha_{\vec{k}} \cos(\omega_0t')) dt'
                \\
    &= 2\omega_{\vec{k}} t + 2 \frac{a_0 \alpha_{\vec{k}}}{\omega_0}
                \sin(\omega_0 t)
                \label{eq:definition_h}
\end{align}
\end{subequations}
Expressing $w_{\vec k}$ by the imaginary part of $z\kk$
 we find the solution for $u_{\vec{k}}(t)$ by integrating Eq.~\eqref{eq:dgl_w}
\begin{equation}
  u_{\vec{k}}(t) = 2a_0 \beta_{\vec{k}}
        \int_0^t \cos(\omega_0 t')\Im z_{\vec k} dt'.
        \label{eq:solution_u}
\end{equation}

In order to focus on the essential slow evolution of $u\kk$
we average out the fast oscillation in $u_{\vec{k}}(t)$ with frequency
$2\omega_0$ and higher. To this end, we replace
$\cos(\omega_0 t) e^{-ih_{\vec{k}}(t)}$ in Eq.~\eqref{eq:solution_z}
by its average over one period $T_0= {2\pi/\omega_0}$.

First, we consider the \textbf{resonant case} $\omega_0=2\omega\kk$. Then
the relation
\begin{equation}
    \phi^{-1} J_1(\phi) = \frac{1}{T_0} \int_0^{T_0}
                \cos(\omega_0 t') e^{-i\omega_0 t' -i\phi \sin(\omega_0 t')} dt'
    \label{eq:simplification_average}
\end{equation}
with $\phi = 2a_0 \alpha_{\vec{k}}/\omega_0$ is directly applicable.
The replacement \eqref{eq:simplification_average} is well justified
if the oscillations at $\omega_0$ are much faster than the slow evolution.
This is fulfilled if oscillation frequency  in the THZ range or smaller
are applied.

We obtain
\begin{equation}
  z_{\vec{k}}(t) = i \frac{\beta_{\vec{k}} \omega_0}{\alpha_{\vec{k}}}
        J_1(\phi) e^{i_{\vec{k}}h_{\vec{k}}(t)} \int_0^t \left(u_{\vec{k}}(t') + 1/2\right)
dt'
   \label{eq:solution_z2}
\end{equation}
with the Bessel function of the first kind $J_1(x)$.
Inserting this result into Eq.~\eqref{eq:solution_u} and using
the averaging of Eq.~\eqref{eq:simplification_average}
in its complex conjugated form again yields
\begin{subequations}
\begin{align}
u\kk(t) &= \Gamma^2 \Re \int_0^t \int_0^{t'} (u\kk(t'')+1/2)dt'' dt'
\\
\Gamma &:= \left(\frac{\beta_{\vec{k}} \omega_0}{\alpha_{\vec{k}}}\right) J_1(\phi).
\end{align}
\end{subequations}
Differentiating twice with respect to time leads to the simple differential equation
\begin{equation}
 u''\kk(t)= \Gamma^2 \left(u_{\vec{k}}(t) + 1/2\right).
\end{equation}
For the relevant initial conditions $u\kk(t=0)=0$ and
$u'\kk(t=0)$ this is solved in a standard way
yielding the result \eqref{eq:tuned} in the main text
as can be easily checked.

Second, we consider the off-resonant case with some detuning
$\delta := 2\omega\kk-\omega_0$.
We assume this detuning to be small so that the oscillations
associated with it can be considered to be small $|\delta| \ll \omega_0$.
Then we apply almost the same steps as before. Using the replacement
\eqref{eq:simplification_average} for $\omega_0$ we arrive at
\begin{equation}
  z_{\vec{k}}(t) = i \frac{\beta_{\vec{k}} \omega_0}{\alpha_{\vec{k}}}
        J_1(\phi) e^{i_{\vec{k}}h_{\vec{k}}(t)} \int_0^t\!\!
        \left(u_{\vec{k}}(t') + 1/2 \right) e^{-i\delta t'} dt' .
        \label{eq:solution_z2_b}
\end{equation}
Then we insert this expression into \eqref{eq:solution_z2}, use the averaging
\eqref{eq:simplification_average} a second time, and obtain
\begin{equation}
  u_{\vec{k}}(t) = \Gamma^2 \Re
        \int_0^t\!\! e^{i\delta t'}\int_0^{t'}\!\! e^{-i\delta t''}
        (u_{\vec{k}}(t'')+1/2) dt'' dt'.
  \label{eq:u-detuned}
\end{equation}
Due to the factors $e^{i\delta t}$ and the real part taken on the right hand side
this is not converted to a closed differential equation by double differentiation.
But differentiating three times yields
\begin{equation}
u'''\kk(t) = (\Gamma^2 -\delta^2)u'\kk(t).
\end{equation}
The initial conditions are $u\kk(t=0)=0$, $u'\kk(t=0)$, and $u''\kk(t=0)=\Gamma^2/2$.
The solution depends on the sign of $\Gamma^2 -\delta^2$
and is given in the main text in \eqref{eq:detuned}.

For constant driving ($\gamma=0$), Fig.~\ref{fig:specific_mode}(a)
displays the magnon occupation $u_{\vec{k}}(t)$ for resonant (blue curve)
and off-resonant driving of type (B) (red curve) as they are
obtained by integrating the Eqs.~\eqref{eq:eom_uvw}
for $\omega_{\vec{k}} = \SI{50}{\per\pico\second}$.
The numerical solution contains fast oscillations at frequency
$2\omega_0$
 with very small amplitudes.
But the slow evolution of $u_{\vec{k}}(t)$ is well captured
by Eqs.\ \eqref{eq:tuned} and \eqref{eq:detuned}.
To illustrate this point, we plot a zoom of the data of
Fig.~\ref{fig:specific_mode}(a) in Fig.~\ref{fig:specific_mode}(b):
the numerical solution wiggles fast around
the analytic result as expected from the above derivation.

\begin{figure}[tb]
  \centering
  \includegraphics[width=1.08\columnwidth]{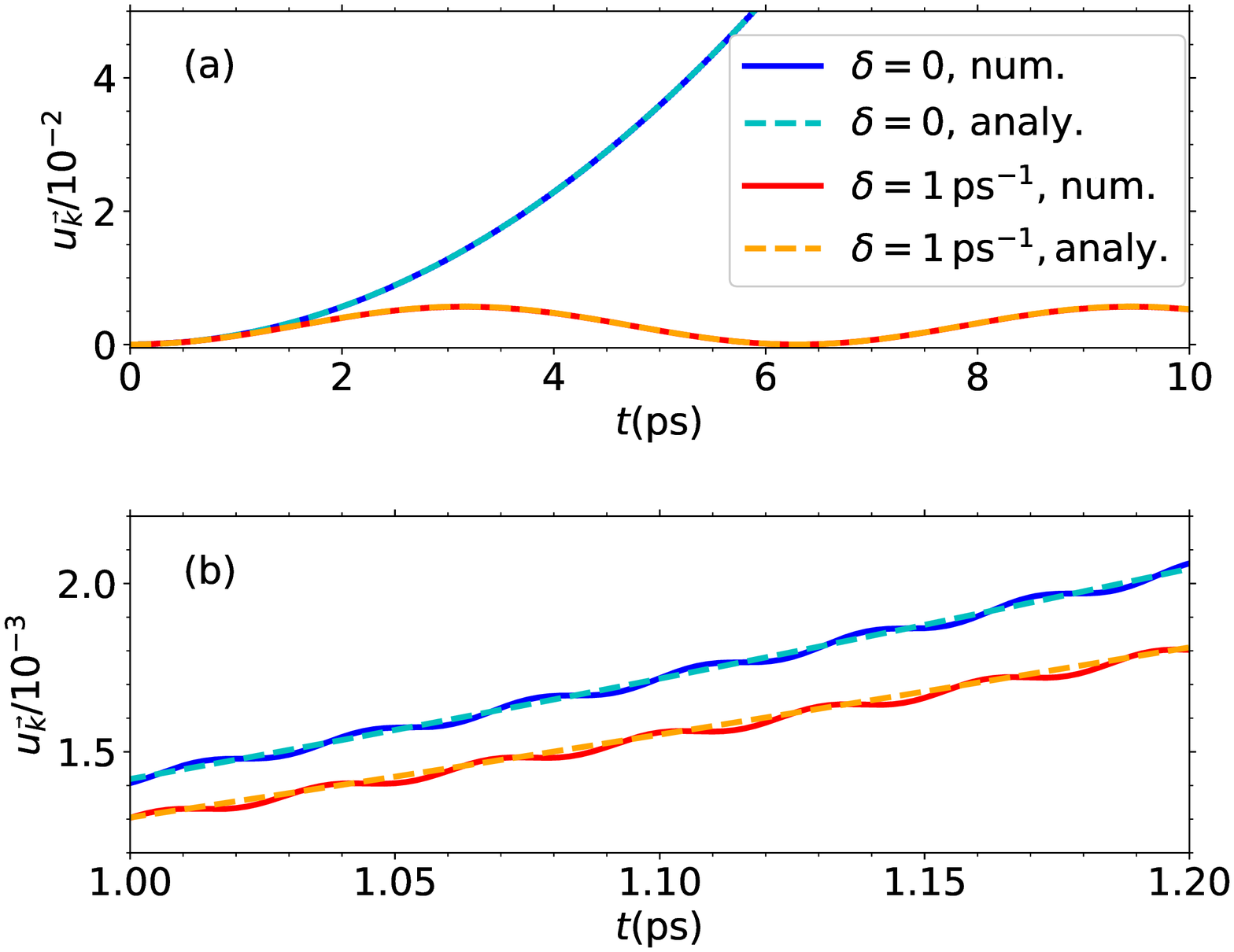}
  \caption{The magnon occupation $u_{\vec{k}}(t)$ of a specific
        magnon mode $\omega_{\vec{k}} = \SI{50}{\per\pico\second}$
  for resonant (dark and light blue) and off-resonant (red and orange)
        driving $\delta=\SI{1}{\per\pico\second}$ with constant amplitude.
        The solid curves represent the full numerical solution while the
        dashed curves represent the analytical approximations
        Eqs.~\eqref{eq:u_res} and \eqref{eq:u_offresb} with
  $\kappa = \num{0.01}, \Gamma= \SI{0.0753}{\per\pico\second},
        \Gamma'= \SI{0.9972}{\per\pico\second}$.
        Panel (a): full time interval; panel (b): zoom of panel (a)
        to show the fast oscillations.
}
  \label{fig:specific_mode}
\end{figure}

For small detuning, $\Gamma > |\delta|$ (not shown),
we obtain a monotonically increasing magnon occupation,
very similar to the completely resonant case. This is important
for the possibility to approximate the integral over the Brillouin zone
by a finite sum over a discrete mesh of eigenfrequencies
because the discretization generically prevents perfect resonance.
For larger detuning ($\Gamma<|\delta|$), the magnon occupation displays
the slow oscillations with the renormalized detuning
$\Gamma' < \delta$ as given by Eq.~\eqref{eq:u_offresb}.

\subsection{Driving by a damped term $a(t)$}
\label{app:damped}

Here we consider the driving term
\be
a(t) = a_0 \cos(\omega_0 t)\exp(-\gamma t),
\ee
Still, \eqref{eq:solution_z} holds if we adapt the definition of $h\kk$ according to
\begin{subequations}
\begin{align}
h_{\vec{k}}(t) &:=
2 \int_0^t (\omega_{\vec{k}} + a_0 \alpha_{\vec{k}}  \cos(\omega_0t')\exp(-\gamma
t)) dt'
                \\
    &= 2\omega_{\vec{k}} t \nonumber
                \\
                & +2 \frac{a_0 \alpha_{\vec{k}}}{\omega_0^2+\gamma^2}
                [e^{-\gamma t}(\omega_0 \sin(\omega_0 t)- \gamma \cos(\omega_0 t)) + \gamma]
                \\
                &\approx 2\omega_{\vec{k}} t .
                \label{eq:definition2_h_c}
\end{align}
\end{subequations}
Unfortunately, the modified averaging similar to  \eqref{eq:simplification_average}
can only be performed in the limit of weak driving $a_0\to0$ where $\Gamma=a_0\beta_k$.
Otherwise, a time dependence remains in $\phi \propto \exp(-\gamma t)$.
But most calculations are indeed done for
weak driving and to understand the effects of these pulses qualitatively the
limit of weak driving is relevant. So we actually use the approximation
Eq.\ \eqref{eq:definition2_h_c} and the averaging
\begin{equation}
1/2 = \frac{1}{T_0} \int_0^{T_0}                 \cos(\omega_0 t') e^{-i\omega_0 t'} dt'
                \label{eq:average2}
\end{equation}
to arrive from \eqref{eq:solution_z} at
\be
 z_{\vec{k}}(t) = i \Gamma
 e^{i_{\vec{k}}h_{\vec{k}}(t)} \int_0^t e^{-\gamma t'}\left(u_{\vec{k}}(t') +
1/2\right) dt'.
\ee
We insert this result in
\be
u_{\vec{k}}(t) = 2\Gamma
        \int_0^t \cos(\omega_0 t')e^{-\gamma t} \Im z_{\vec k} dt',
\ee
apply the averaging a second time, and finally obtain
\be
\label{eq:integration_tuned}
u\kk(t) = \Gamma^2 \int_0^t dt' e^{-\gamma t'}\int_0^{t'} dt'' e^{-\gamma t''}
(u\kk(t'') + 1/2).
\ee
At this stage, one can proceed in two ways. One is to focus
on low occupations $2u\kk\ll1$ so that $u\kk$ in the integrand
can be neglected. Double integration straightforwardly yields
 Eq.\ \eqref{eq:tuned_damped} given in the main text.
Alternatively, twice differentiating
of Eq.\ \eqref{eq:integration_tuned}
yields the differential equation
\be
u\kk''= \Gamma^2 e^{-2\gamma t} (u\kk+1/2) -\gamma u\kk'
\ee
which needs to be solved for the initial conditions $u\kk(0)=0=u\kk'(0)$.
No closed expression can be found for its solution, but it
is easily integrated by any computer algebra program.

Similarly, we can tackle the detuned case yielding
\be
\label{eq:integration_detuned}
u\kk(t) = \Gamma^2\Re \int_0^t\!\! dt' e^{(i\delta-\gamma) t'}\!\!\int_0^{t'}\!\! dt''
e^{-(i\delta+\gamma) t''} \left(u\kk(t'') + \frac{1}{2}\right).
\ee
Again, one can proceed in two ways. One is to focus
on low occupations $2u\kk\ll1$ so that $u\kk$ in the integrand
can be neglected. Double integration straightforwardly yields
Eq.\ \eqref{eq:detuned_damped} given in the main text.
Alternatively, a differentiation  of Eq.\ \eqref{eq:integration_detuned},
multiplication with $\exp(\gamma t)$ and two more
differentiation yield the differential equation
\begin{align}
\nonumber
u\kk''' &= \left(\Gamma^2 e^{-2\gamma t} -\delta^2-\gamma^2\right)
u\kk'
\\
&\qquad\quad -2\gamma u\kk'' - \gamma\Gamma^2 e^{-2\gamma t} (u\kk(t)+1/2).
\end{align}
This equation can easily be solved numerically with the initial conditions
$u\kk(0)=0=u\kk'(0)$ and $u\kk''(0)=\Gamma^2/2$.

\subsection{Driving by a damped
term $a(t)$ in presence of magnon relaxation}
\label{app:relax}

Here we extend the previous arguments to the case where the
phenomenological relaxation of magnetic modes is included.
That means that $\gamma_r>0$ is considered and we start from
the general Eqs.\ \eqref{eq:eom_uvw} and \eqref{eq:eom_z}.
Integrating $u\kk$ yields
\be
\label{eq:u_full}
u\kk(t) = 2\Gamma e^{-\gamma_r t} \int_0^t \cos(\omega_0t') e^{(\gamma_r-\gamma)t'}
\Im z\kk(t') dt'.
\ee
The integration of \eqref{eq:eom_z} yields
\be
\label{eq:z_full}
z\kk(t) = i e^{ih\kk-\gamma_rt} \int_0^t f\kk(t') e^{-ih\kk+\gamma_rt'}dt'.
\ee
As before we average the quickly oscillating part in \eqref{eq:z_full}
using \eqref{eq:average2} and insert the result into \eqref{eq:u_full}.
Then we average a second time using \eqref{eq:average2} to obtain
\be
\label{eq:u_result_tuned}
u\kk(t) = \Gamma^2 e^{-\gamma_r t}  \int_0^t\!\! dt' e^{-\gamma t'}
 \int_0^{t'} \!\! dt'' e^{-\gamma t''}(u\kk + e^{\gamma_r t''}/2)
\ee
in the resonant case, i.e., for $\omega_0=2\omega\kk$.
For deriving the corresponding differential equation it is convenient to define
\be
\tilde u\kk(t) := \exp(\gamma_r t) u\kk(t).
\ee
Differentiating $\tilde u\kk(t)$ twice
yields the differential equation
\be
\tilde u''\kk(t) = -\gamma \tilde u'\kk +\Gamma^2 e^{-2\gamma t}
\left(e^{-\gamma_r t} \tilde u\kk+  e^{\gamma_r t}/2 \right),
\ee
which is to be solved for $\tilde u\kk(0) =0 =\tilde u'\kk(0)$.

Alternatively, we restrict ourselves to small values of $u\kk$ and
directly integrate \eqref{eq:u_result_tuned} providing us with
\be
u\kk(t) =\frac{\Gamma^2}{2}
\frac{(\gamma-\gamma_r)e^{-\gamma_r t} +\gamma e^{-2\gamma t} -(2\gamma-\gamma_r)
e^{-(\gamma+\gamma_r) t}}{\gamma(\gamma-\gamma_r)(2\gamma-\gamma_r)}.
\ee
Note that this leads to an initial rise followed by a maximum and subsequent
bi-exponential decay governed by $e^{-\gamma_r t}$ and $e^{-\gamma t}$.

Finally, we address the detuned case for which go again through the same
steps as above. Instead of Eq.\ \eqref{eq:u_result_tuned} we obtain
\begin{widetext}
\begin{align}
\label{eq:u_result_detuned}
u\kk(t) &= \Gamma^2 e^{-\gamma_r t}\Re  \int_0^t dt' e^{(i\delta-\gamma) t'}
 \int_0^{t'}  dt'' e^{-(i\delta+\gamma) t''}(u\kk + e^{\gamma_r t''}/2).
\end{align}
Appropriate triple differentiation allows to derive the differential equation
\begin{eqnarray}
\tilde u'''\kk(t) &=& -2\gamma \tilde u''\kk -
\left(\delta^2+\gamma^2-\Gamma^2e^{-(2\gamma+\gamma_r)t}\right)\tilde u'\kk
-(\gamma+\gamma_r)\Gamma^2 e^{-(2\gamma+\gamma_r)t} \tilde u\kk
- (\gamma-\gamma_r)\Gamma^2 e^{(-2\gamma+\gamma_r)t}/2
\end{eqnarray}
which needs to be solved for the initial conditions
$\tilde u\kk(0) =0 =\tilde u'\kk(0)$ and $\tilde u''\kk(0) =\Gamma^2/2$.

For small values of $u\kk$ we can integrate \eqref{eq:u_result_detuned}
to obtain
\begin{eqnarray}
u\kk(t) &=& \frac{\Gamma^2}{2} \Bigg( \frac{(\gamma-\gamma_r)e^{-2\gamma
t}}{(2\gamma-\gamma_r)
(\delta^2+(\gamma-\gamma_r)^2)}
-\frac{[(\delta^2 -\gamma\gamma_r+\gamma^2)\cos(\delta t)+
\delta\gamma_r\sin(\delta t)]e^{-(\gamma+\gamma_r) t}}
{(\delta^2+\gamma^2)(\delta^2+(\gamma-\gamma_r)^2)}
\nonumber \\
&&
 + \frac{\gamma e^{-\gamma_r t}}{(2\gamma-\gamma_r)
(\delta^2+\gamma^2)}
\Bigg).
\end{eqnarray}

\end{widetext}

\end{document}